\documentclass[aps,prd,showpacs,nofootinbib,floats,floatfix,preprintnumbers,groupedaddress,twocolumn]{revtex4}

\usepackage{bm}
\usepackage{latexsym}
\usepackage{dcolumn}
\usepackage{amsmath,amsfonts,amssymb}
\usepackage{graphicx,epsfig}
\usepackage{amsmath}
\usepackage{fancyhdr}
\usepackage{hyperref}
\usepackage{graphicx,epstopdf}
\usepackage{physics}
\usepackage[utf8]{inputenc}

\hypersetup{
	colorlinks   = true, 
	urlcolor     = blue, 
	linkcolor    = blue, 
	citecolor   = red 
}

\def\be{\begin{equation}}
	\def\ee{\end{equation}}
\def\beq{\begin{eqnarray}}
	\def\eeq{\end{eqnarray}}

\newcommand{\levicivita}{}
\def\levicivita#1#{\tensor#1{\epsilon}}
\newcommand{\inner}[3]{\langle #1\vert#2\vert#3\rangle}
\newcommand{\Trf}[1]{\text{Tr}_{f}}
\newcommand{\avg}[4]{\langle\Phi_{#1}(#2_{#1})\Phi_{#3}(#4_{#3})\rangle}

\begin{document}
	\title{Entangled quantum Unruh Otto engine is more efficient}
	\author{Gaurang Ramakant Kane\footnote {\color{blue} gsrkane@gmail.com}}
	\author{Bibhas Ranjan Majhi\footnote {\color{blue} bibhas.majhi@iitg.ac.in}}
	
	\affiliation{Department of Physics, Indian Institute of Technology Guwahati, Guwahati 781039, Assam, India}
	
	\date{\today}
	
	\begin{abstract}
		We propose a relativistic quantum Otto cycle between an entangled state of two qubits and their composite excited (or ground) state whose efficiency can be greater than the usual single qubit quantum Otto engine. The hot and cold reservoirs are constructed by providing uniform accelerations to these qubits along with the interaction between the background field and individual qubits. The efficiency, as measured from one of the qubits' frame, not only depends on the energy gap of the states but also the relative acceleration between them. For lower acceleration of our observer's qubit compared to the other one, the cycle is more efficient than the single qubit quantum Otto engine. Furthermore, a complete protocol to construct such a cycle is being provided.
	\end{abstract}
	
	
	\maketitle
	
		The combination of quantum mechanics (QM) and relativity (appropriately defined through the language of quantum field theory (QFT)) brings many surprising results for us, like Hawking \cite{Hawking:1974rv,Hawking:1974sw} and Unruh \cite{Unruh:1976db,Unruh:1983ms} effects and it is expected to bring more in the future. Along with the efforts of combining QM and relativity, continuous efforts have been made to study the  foundations of the thermodynamics of the quantum systems \cite{Book1}. If a quantum system, (e.g. qubits which we will be considering here) has Hamiltonian $H_0(t)$ and is described by density operator $\rho(t)$, the energy at any instant is determined by the expectation value $\langle E(t)\rangle = \textrm{Tr}[\rho(t)H_0(t)]$. The variation of it from time $t_i$ to $t_f$ yields the definitions of heat transfer and work done on the system as \cite{Book1}
	\begin{equation}
		\langle Q\rangle=\int_{t_{i}}^{t_{f}}\text{Tr}\Big(\dfrac{d\rho(t)}{dt}H_0\Big)dt~,
		\label{B3}
	\end{equation}
	and
	\begin{equation}
		\langle W\rangle=\int_{t_{i}}^{t_{f}}\text{Tr}\Big(\rho(t)\dfrac{dH_0}{dt}\Big)dt~,
		\label{B4}
	\end{equation}
	respectively.
	Within the area of thermodynamics of quantum systems, intense focus is devoted to quantum heat engines. Although the concept was introduced in the year $1959$ \cite{PRL1}, lot of attention and interest have imparted after works by Kieu \cite{PRL2,EPJD1} who proposed the single-qubit {\it quantum Otto engine} (QOE). Similar to the {\it classical Otto engine} (COE), QOE consists of two adiabatic and two isochoric stages. The efficiency of this engine is given by 	
	\begin{equation}
		\eta_0 = 1- \frac{\omega_1}{\omega_2}~.
		\label{B5}
	\end{equation}
In contrast to COE, production of finite work in QOE is possible provided we have the condition $\omega_1T_H>\omega_2T_C$, where $T_H$ $(T_C)$ is the temperature of hot (cold) reservoir and $\omega_1$ $(\omega_2)$ is the energy gap before (after) adiabatic process.  
	
	Recently in a relativistic setup, a QOE has been introduced \cite{Arias:2017kos} in the light of the Unruh effect (called as {\it Unruh quantum Otto engine} (UQOE)). In the adiabatic stages, the qubit is moving with a constant velocity. The hot and cold reservoirs for the other two stages of the cyclic process are being mimicked by allowing this qubit to uniformly accelerate in addition to the interaction with the background massless real scalar field. The point-like qubit is taken to be a monopole. Its behaviour is identical to QOE. The efficiency is again $\eta_0$, and the condition for producing work also comes out to be similar. The same analysis has been extended in \cite{Gray:2018ifq} for different background quantum fields and interactions. Although the efficiency remains unchanged, the work output comes out to be different for different background fields and interactions. The UQOE has also been constructed between the degenerate states \cite{Xu:2019hea}.
	Another important phenomenon in QM is the quantum entanglement. People have found that the entanglement between two relativistic accelerating detectors has a profound impact on the Unruh effect \cite{Arias:2015moa,Costa:2020aqa,Rodriguez-Camargo:2016fbq,Barman:2021oum}. On the other side, the entanglement between two states is relativistically observer-dependent \cite{Alsing:2003cy,FuentesSchuller:2004xp,Downes:2010tv} -- an accelerated observer will see the degradation of entanglement. A similar conclusion has also been drawn with localized field modes by trapping them within a cavity \cite{Additional4}. Relativistic effects on quantum teleportation with respect to a non-uniformly moving frame have been investigated as well, and it has been observed that the upper bound to the optimal fidelity is degraded due to this motion \cite{Additional3}. The equivalence principle lays a profound path for understanding several features of gravity through uniformly accelerated frames. Therefore, understanding the accelerated frame can enlighten few aspects of gravity. Nevertheless, a direct encounter of gravity has also been investigated in few instances. For example, it has been shown that the gravity induces decoherence of stationary matter superposition states \cite{Additional1,Additional2} (also see \cite{Nesterov:2020exl}). Moreover, entanglement  harvesting between two causally disconnected detectors is possible under a relativistic setup \cite{Reznik:2002fz,Sachs:2017exo,Koga:2018the,Ng:2018ilp,Koga:2019fqh,Barman:2021bbw}. Within this relativistic setup, harvesting of coherence in the initial state of the scalar field is possible as well \cite{Kollas:2020ghd,Kollas:2021nqy}. These findings provide robust evidences of new understandings of our physical systems when combining quantum mechanics with relativity in terms of QFT language.
	
	Here, combining entanglement and relativity, we propose an Otto engine composed of two qubits. We construct the Otto cycle between their composite entangled and the excited (or ground) states. The steps of the cycle are similar to those of UQOE (we refer \cite{Arias:2017kos} for details of the four stages of UQOE cycle), and therefore we name it as {\it entangled Unruh quantum Otto engine} (EUQOE). The description of these stages is discussed later. Surprisingly, the efficiency $\eta_E$, measured from the frame of one of the qubits, can be larger than $\eta_0$ by tuning the accelerations of the two qubits. If the engine is initially in the maximally entangled state and our frame of reference is not accelerating, then $\eta_E=2\eta_0$. We find that this enhancement arises as a result of both entanglement and relativity. A protocol for building a EUQOE with $\eta_E>\eta_0$ is being provided at the end. We also show in our protocol that in any case, $\eta_E$ is less than unity.


	Our system is composed of double two-level identical qubits (sometimes we call them detectors). Their proper times are denoted by $\tau_{1}$ and $\tau_{2}$, respectively. 	
	The total internal free Hamiltonian of the detectors, following \cite{Dicke:1954zz,Arias:2015moa}, is chosen to be
	\begin{equation}
		H_0 = \frac{\omega(\tau)}{2}\Big(\dfrac{d\tau_{1}}{d\tau}S_{1}^{z}\otimes\mathbb{I}_{2}+\dfrac{d\tau_{2}}{d\tau}\mathbb{I}_{1}\otimes S_{2}^{z}\Big)~,
		\label{eqn:internalhamiltonian}
	\end{equation}
	where $S_{i}^{z} = (\ket{e_j}\bra{e_j} - \ket{g_j}\bra{g_j})$ with $\ket{g_j}$ and $\ket{e_j}$ represent ground and excited states of $j^{th}$ detector, respectively. $\omega$ is the energy difference between the levels of each qubit. Here, $\ket{g_i}$ has energy $-\omega/2$ while $\ket{e_i}$ corresponds to energy eigenvalue $\omega/2$ and $\mathbb{I}_{j}$ denotes identity operator for $j^{th}$ detector. $\tau$ denotes the proper time of the detector from which all the measurements will be done. In other words, $\tau$ is the proper time for the observer, who is analyzing the process. For our present purpose, we choose this to be $\tau=\tau_1$ as everything will be measured from the first detector's frame.  Hence, we will substitute $d\tau_1/d\tau =1$. On the other hand $\tau_1$ and $\tau_2$ are not the same when the two detectors move with different velocities or accelerations; i.e. in a general case $d\tau_2/d\tau \neq 1$. They are shown to be related through the relation $\tau_2=\alpha\tau_1$, where $\alpha$ is the Lorentz factor corresponding to the relative velocity between them during their constant velocity motions while $\alpha$ is the ratio of accelerations of the detectors during their uniform accelerated motions (see \cite{Rodriguez-Camargo:2016fbq} for details. We have also presented an analysis in Supplementary material \cite{App1B}). If the qubits are moving with different velocities or accelerations then $\alpha\neq 1$. 
	
	Here, the two-qubit system has the following four normalised eigenstates: 
	\begin{eqnarray}
		E_{e} &=& \omega~,~~|e\rangle = |e_{1}\rangle |e_{2}\rangle~;\nonumber\\
		E_{s} &=& 0~,~~|s\rangle = \frac{1}{\sqrt{2}}(|e_{1}\rangle 
		|g_{2}\rangle+|g_{1}\rangle |e_{2}\rangle)~;\nonumber\\
		E_{a} &=& 0~,~~|a\rangle = \frac{1}{\sqrt{2}}(|e_{1}\rangle 
		|g_{2}\rangle-|g_{1}\rangle |e_{2}\rangle)~;\nonumber\\
		E_{g} &=& -\omega~,~~|g\rangle = |g_{1}\rangle |g_{2}\rangle~.
		\label{B6}
	\end{eqnarray}
	We consider each detector as a monopole and it interacts with the background massless real scalar field through linear coupling of the form $\mu_{i}\kappa(\tau_{i})m_{i}\Phi(X_{i}(\tau_{i}))$ (i.e. Unruh De-Witt type) \cite{BookBD}. Here $\Phi(X)$ is the \emph{massless real scalar field} and $\kappa(\tau_{i})$ denotes the switching function which decides whether the interaction is on or off. This is unity during the interaction and vanishes otherwise. $m_{i}$ represents the monopole operator for the $i^{th} detector$, given by
	\begin{equation}
		m_{i}=\ket{g_{i}}\bra{e_{i}}+\ket{e_{i}}\bra{g_{i}}	~,	
		\label{B9}
	\end{equation}
	and $\tau_i$ is the proper time of this detector. If our observer's proper time is $\tau$ then the interaction term for the evolution process can be represented as $\int d\tau_i \mu_{i}\kappa(\tau_{i})m_{i}\Phi(X_{i}(\tau_{i}))$ where, since $\tau_i$ is a function of $\tau$, we have to use $d\tau_i = (d\tau_i/d\tau) d\tau$.
	Therefore, the total interaction Hamiltonian for the two qubits-massless real scalar field system is chosen as
	\begin{equation}
		H_{int}=\sum_{i=1,2}\mu_{i}\kappa(\tau_{i})m_{i}\Phi(X_{i}(\tau_{i}))\dfrac{d\tau_{i}}{d\tau}~.
		\label{B7}
	\end{equation}
	Similar argument has been considered in \cite{Tjoa:2020eqh,Gallock-Yoshimura:2021yok} for the derivative coupling case (for monopole coupling cases see \cite{Koga:2018the,Ng:2018ilp,Barman:2021oum,Barman:2021bbw}).
	As the detectors are identical, we keep the coupling constants as $\mu_{1}=\mu_{2}=\mu$. The above introduction of the scalar field $\Phi$ brings the role of QFT in our analysis, which we shall encounter later.
	
		This two-qubits system has four eigenstates (see Eq. (\ref{B6})). So several Otto cycles can be constructed between any of the two eigenstates. It is already known from earlier analysis \cite{Barman:2021oum} that if the detectors are in uniform acceleration then the transition only from symmetric (i.e. $\ket{s}$) or antisymmetric (i.e. $\ket{a}$) to excited (i.e. $\ket{e}$) or ground (i.e. $\ket{g}$) states is possible whereas from $\ket{g}$ to $\ket{e}$ is not possible. Moreover, the transition rate for $\ket{s}\to\ket{g}$ is same as that of $\ket{s}\to\ket{e}$. Similarly, transition rates between $\ket{a}\to\ket{g}$ and $\ket{a}\to\ket{e}$ are also equal. Motivated by these facts we choose our Otto cycle between $\ket{s}$ and $\ket{e}$ (or $\ket{g}$). The other choice can be between $\ket{a}$ and $\ket{e}$ (or $\ket{g}$). Note that $\ket{s}$ and $\ket{a}$ are maximally entangled states. To keep our discussion general, we choose our Otto cycle between 
	the entangled state
	\begin{equation}
		\ket{\chi}=b_{1}\ket{e_{1}}\ket{g_{2}}+b_{2}\ket{g_{1}}\ket{e_{2}}~,
		\label{B10}
	\end{equation}
	and  $\ket{e}$ (consideration of $\ket{g}$ does not change the final conclusion). The coefficients $b_{1}$ and $b_{2}$ satisfy
	$\vert b_{1}\vert^{2}+\vert b_{2}\vert^{2}=1$ in order to make $\ket{\chi}$ a normalised state. Here the values of $b_1$ and $b_2$ can be chosen to be $b_1=1/\sqrt{2}=b_2$ for $\ket{s}$  and $b_1=1/\sqrt{2}, b_2 = -1/\sqrt{2}$ for $\ket{a}$. For the moment we keep them in general form and at the end we will consider these specific values. Therefore our initial density matrix  for the two-qubits system is 
	\begin{equation}
		\rho_{A_0}=p\ket{e}\bra{e}+q\ket{\chi}\bra{\chi}~,		
		\label{B11}
	\end{equation}
	where, $q=1-p$. Initially, the energy gap between $\ket{e}$ and $\ket{\chi}$ is $\omega=\omega_1$. For the two-qubits plus the scalar field system we choose the density matrix as $\rho_0^I = \rho_{A_0}\otimes\rho_{f_0}$, where $\rho_{f_0} = \ket{0_M}\bra{0_M}$ by considering the field initially is in Minkowski vacuum $\ket{0_M}$. In our analysis, since we are interested only to qubits system, all the relevant quantities will be evaluated by taking trace over field states.
	
	Let us now describe the four stages of our EUQOE. 
	As mentioned earlier, we aim to calculate work and heat in the frame of detector 1. For this, we require the duration of interaction in terms of proper time $\tau_{1}$. This duration is calculated by analysing the kinematics of the entire process in reference to a fixed inertial frame associated with Minkowski coordinates $(T,X)$.
	The steps in the cycle are similar to those proposed in UQOE \cite{Arias:2017kos} for the single-qubit system.
	(1) The first phase is the adiabatic phase, where no heat is exchanged with the environment. In this phase, two detectors move with the constant velocities $-v_1$ and $-v_2$, respectively, and the switching function in Eq. (\ref{B7}) is chosen to be $\kappa(\tau_i) = 0$. The velocities are measured with reference to the fixed inertial frame as mentioned above. The time intervals for this stage are $\Delta\tau^1_1$ and $\Delta\tau^1_2$. In this process, $\omega_{1}$ changes to $\omega_{2}$ at the cost of work being done on the system. (2) Once the energy difference changes to $\omega_{2}$, the qubits start to accelerate uniformly with proper accelerations ${a_H}_1$ and ${a_H}_2$, respectively. In this stage, the interaction with the background scalar field is on; i.e. $\kappa(\tau_i) = 1$. This process continues till they reach the velocities $+v_1$ and $+v_2$ from $-v_1$ and $-v_2$. The time duration of this stage with respect to individual detectors are $\Delta\tau^2_1$ and $\Delta\tau^2_2$, respectively. The general relations between the Minkowski coordinates and the proper time of a uniformly accelerated detector, moving along the Minkowski $X$-axis, are
	\begin{equation}
		T = \frac{1}{a}\sinh(a\tau); \,\,\,\ X = \frac{1}{a}\cosh(a\tau)~.
		\label{B12}
	\end{equation}
	From (\ref{B12}) we find the proper time $\tau$ at which the velocity is $v$ as
	$v= \tanh(a\tau)$. Therefore the time at which the velocity is $-v$ is $-\tau$. So the total duration for changing from $-v$ to $v$ is $2\tau$. This yields
	\begin{equation}
		\Delta\tau^2_1 = \frac{2}{{a_H}_1}{\textrm{arctanh}}(v_1); \,\,\,\ \Delta\tau^2_2 = \frac{2}{{a_H}_2}{\textrm{arctanh}}(v_2)~.
		\label{B14}
	\end{equation}
	The relation between these time intervals during the accelerated stage, following \cite{App1B}, is 
	\begin{equation}
		{a_H}_1\Delta\tau^2_{1}={a_H}_{2}\Delta\tau^2_{2}~.
		\label{B15}
	\end{equation}
	This implies that we can take the velocities of both the detectors to be same; i.e. $v_1= v_2=v$ (say). Therefore in stage 1 they are moving with the same velocity $-v$ and at the end of stage 2 their velocity changes to $+v$. Since in this stage they are at different proper accelerations, the durations of interaction with the background scalar field are different. Due to their accelerations, the interaction with scalar fields yields transition between the energy levels in the individual detectors and each qubit feels the Minkowski vacuum as thermal bath \cite{Unruh:1976db,Unruh:1983ms}. Our observer (the frame of first qubit) will then see himself in a thermal bath and then the Otto engine will absorb heat from the environment.
	(3) In this stage, the detectors stop accelerating and they move with constant velocity $+v$. The interaction is also turned off. Now, again this is an adiabatic phase with time durations $\Delta\tau^3_1$ and $\Delta\tau^3_2$, where $\omega_{2}$ changes to $\omega_{1}$. Similar to stage 1, the engine will perform work without any heat exchange with the environment.  (4) Finally, in the last part, similar to the second step, both the detectors decelerate with $a_{C_1}$ and $a_{C_2}$ from velocity $+v$ to $-v$. We again switch on the interaction between the detectors and the scalar field for durations $\Delta\tau^4_1$ and $\Delta\tau^4_2$. The two detectors will see cold reservoirs, and due to similar reasons, as discussed in stage 2, they will reach the same velocity but with different durations. Here our observer will find the background as a cold bath, and hence the Otto engine will reject heat to the environment. For future purpose denote $\alpha_{a_H} = a_{H_1}/a_{H_2}$ and $\alpha_{a_C} = a_{C_1}/a_{C_2}$.

	In step 2, the density matrix will change from $\rho_{A_0}\rightarrow \rho_{A_0}+\delta\rho^{H}$. In step 4, it will change from $\rho_{A_0}+\delta\rho^{H}\rightarrow \rho_{A_0}+\delta\rho^{H}+\delta\rho^{C}$. Hence, for the cyclic process, we impose a constraint:
	\begin{equation}
		\delta\rho^{H}+\delta\rho^{C}=0~.
		\label{B16}
	\end{equation}
	$H$ and $C$ stand for heating and cooling, respectively.
Using the definitions (\ref{B3}) and (\ref{B4}) we will now find the amount of work and heat transfer in each stage. To calculate them, we need to know how the density operator changes with time. The results, till second-order perturbation, are derived using the standard procedure \cite{App2B}.

	
	In the adiabatic process (i.e. in stage $1$ and stage $3$), the work is done on or done by the engine without any heat exchange with the environment. The quantum definition of work is given by (\ref{B4}). The internal free Hamiltonian for the two qubit system is (\ref{eqn:internalhamiltonian}), which for our chosen basis (i.e. $\{\ket{e_1,e_2},\ket{e_1,g_2},\ket{g_1,e_2},\ket{g_1,g_2}\}$), has a matrix representation as follows:
\begin{equation}
	H_{0}=\dfrac{\omega}{2}\begin{pmatrix}
		(1+\alpha)&0&0&0\\
		0&(1-\alpha)&0&0\\
		0&0&(-1+\alpha)&0\\
		0&0&0&(-1-\alpha)\\
	\end{pmatrix} \equiv \omega h_\alpha~.
	\label{B25}		
\end{equation}
$h_\alpha$ denotes the matrix whose explicit form has been obtained through $\tau_2 = \alpha \tau_1$ \cite{App1B} with the choice $\tau=\tau_1$. We put $\alpha$ in the subscript which will be denoted as $\alpha_v$ during the motions with constant velocity while it will be $\alpha_a$ during the motions with uniform accelerations. 
In stages, $1$ and $3$, both the detectors move with a constant velocity and only the separation between the energy levels changes. Whereas, the population of the energy levels remains the same. So $\omega$ is a function of time while other quantities remain fixed.
Hence, work done in stage $1$ is
\begin{equation}
	\langle W\rangle_{1}=(\omega_{2}-\omega_{1})\text{Tr}(\rho_{A_0}h_{\alpha_{v}})~,
	\label{B27}
\end{equation}
and the same in stage $3$ is
\begin{equation}
	\langle W\rangle_{3}=(\omega_{1}-\omega_{2})\text{Tr}\Big((\rho_{A_0}+\delta\rho^{H})h_{\alpha_{v}}\Big)~.
	\label{B28}
\end{equation} 
The total work done on the system is then
\begin{equation}
	\langle W\rangle_{tot} = \langle W\rangle_{1}+\langle W\rangle_{3}=(\omega_{1}-\omega_{2})\textrm{Tr}(\delta\rho^{H}h_{\alpha_{v}})~,
	\label{B29}
\end{equation}
which depends on the change of density matrix during the heating process in stage $2$.


Heat exchange between the system and the environment takes place in stages $2$ and $4$ as there is population change of the energy levels. The quantum heat is defined by (\ref{B3}). Since here
$H_{0}$ is constant of time, no work is done on or by the system. In stage $2$, the amount of heat absorbed by the engine is
\begin{equation}
	\langle Q\rangle_{2}=\omega_{2}\text{Tr}(\delta\rho^{H}h_{\alpha_{a_H}})~,
	\label{B31}
\end{equation}
and the heat transferred to the environment in stage $4$ turns out to be
\begin{equation}
	\langle Q\rangle_{4}=\omega_{1}\text{Tr}(\delta\rho^{C}h_{\alpha_{a_C}}) = -\omega_{1}\text{Tr}(\delta\rho^{H}h_{\alpha_{a_C}})~.
	\label{B32}
\end{equation}
In the last step (\ref{B16}) has been used.
Therefore net heat absorbed by our engine is
\begin{equation}
	\langle Q \rangle_{tot} = \langle Q\rangle_{2} + \langle Q\rangle_{4} = \omega_2  \text{Tr}(\delta\rho^{H} h_{\alpha_{a_H}}) - \omega_1\text{Tr}(\delta\rho^{H}h_{\alpha_{a_C}})~.
	\label{B33}
\end{equation}

An important difference between the single-qubit system and our present two-qubits system is as follows. Usually, for the cyclic process the condition to be imposed is given by (\ref{B16}). Note that it appears both in single-qubit and in our system. But in present engine we have additional constraint to be imposed in order to maintain the conservation of energy in the cyclic process; i.e. $\langle W\rangle_{tot} + \langle Q \rangle_{tot}=0$. This is trivially satisfied in single qubit system, but here we need to impose the following relation 
\begin{equation}
	\omega_{2}\text{Tr}(\delta\rho^{H}h_{\alpha_{a_H}})-\omega_{1}\text{Tr}(\delta\rho^{H}h_{\alpha_{a_C}})=(\omega_{2}-\omega_{1})\text{Tr}(\delta\rho^{H}h_{\alpha_{v}})~.
	\label{B34}
\end{equation}
We will comeback to these conditions later.

The efficiency of a cyclic system is the ratio between the total work done by the system and heat absorbed by the same system; i.e. $\eta_E = -\langle W \rangle_{tot}/\langle Q \rangle_2$. Therefore, for our Otto cycle it turns out to be
\begin{equation}
	\eta_{E} = \Big(1-\dfrac{\omega_{1}}{\omega_{2}}\Big)\dfrac{\textrm{Tr}(\delta\rho^{H}h_{\alpha_{v}})}{\text{Tr}(\delta\rho^{H}h_{\alpha_{a_H}})} = \eta_0 \dfrac{\textrm{Tr}(\delta\rho^{H}h_{\alpha_{v}})}{\text{Tr}(\delta\rho^{H}h_{\alpha_{a_H}})}~,
	\label{B35}
\end{equation}
where $\eta_0$ is given by (\ref{B5}). 
This is quite a general expression that depends on the chosen states to construct the Otto cycle. 
For our case, this is given by (\ref{B11}). Now we aim to find $\eta_E$ for our specific configuration.

To evaluate (\ref{B35}) it is now sufficient to calculate $\textrm{Tr}(\delta\rho^{\alpha}h_{\alpha'})$ in which at the end we will choose $\alpha'=\alpha_v=1$ and $\alpha' = \alpha_{a_H}$ with $\alpha=\alpha_{a_H}$ to get explicit forms of numerator and denominator. As we are interested only in the two-qubits system, this term appeared after tracing over field states. Therefore, the trace operation in this term (denoted by ``Tr'' here), which is to be evaluated to calculate the efficiency, is over the basis of two-qubits system.
During the heating stage, $\rho$ is changing and velocity changes from $-v$ at $-\tau$ to $v$ at $\tau$. Therefore we choose the interval of $\tau_1$ for interaction as $[-\tau_a,\tau_a]$.  With this we find \cite{App5B} for $\ket{\chi} = \ket{s}$ and $p=0$ (i.e. initially the system is in $\ket{\chi}$ and so $q=1$)
\begin{eqnarray}
	&&\text{Tr}(\delta\rho^{\alpha} h_{\alpha'})=i\mu^{2}\alpha(1+\alpha')[e^{i\omega(\alpha-1)\tau_{a}}+e^{-i\omega(\alpha-1)\tau_{a}}]
	\nonumber
	\\
	&&\times \int_{-\tau_a}^{\tau_a}\int_{-\tau_a}^{\tau_a}\sin(\omega(\alpha\tau_{1}'-\tau_{1}''))G_{12}(\tau_{1}'',\tau_{2}'(\tau_{1}'))d\tau_{1}''d\tau_{1}'~.
	\nonumber
	\\
	\label{B39}
\end{eqnarray}
In the above we denote $G_{ij} (\tau'_i,\tau''_j)= \langle\Phi_i(\tau'_i)\Phi_j(\tau''_j)\rangle$ which is the positive frequency Wightman function for real massless scalar field. The final form, as we are measuring from the frame of first detector, is achieved by expressing second detector's proper time in terms of that of the first one.
Then (\ref{B35}) gives us the measure of efficiency as
\begin{equation}
	\eta_E = \eta_0\Big(\dfrac{2}{1+\alpha_{a_H}}\Big)~.
	\label{B41}
\end{equation}
The same can also be obtained for choosing antisymmetric entangled state as the initial one, i.e. for $\ket{\chi}=\ket{a}$. 

The result (\ref{B41}) implies that the efficiency can be greater than $\eta_0$ if one chooses $\alpha_{a_H}<1$; i.e. if $a_{H_1}<a_{H_2}$; or in other words the second qubit must accelerate more than the first one. It implies that the second detector must be closer to $X=\pm T$ null surface than the other one. By increasing the difference between the accelerations with the above condition, we can have a more efficient Otto engine. Otherwise, the efficiency of our entangled system is less than that of the single-qubit system.
Suppose, in the second stage, the first detector is not accelerating (i.e. it is either at rest or in uniform velocity) while the second one is accelerating. In this case we have $\alpha_{a_H} = (a_{H1}/a_{H2}) = 0$. Note that although $a_{H_1}=0$, still the whole system will see the background as a thermal bath since the other qubit is in acceleration and the initial state is an entangled one. Then the efficiency is given by $\eta_E = 2\eta_0$.

	In the analytical analysis done so far, we have seen that different variables affect the process. They appear in $\delta\rho^{H}$ and hence finally appear in the expression for efficiency. We now aim to provide a protocol that will allow one to find an Otto cycle with $\eta_{E}>\eta_{0}$ for appropriate choices of various variables. We first list the variables we have to investigate and the conditions those are required.
	The variables are $\alpha_{a_H},\tau_{a},\alpha_{a_C}$ and the conditions to satisfy are as follows.
	(i) Cyclic condition (\ref{B16}) along with the energy conservation, given by (\ref{B34}): A {\it sufficient} (not necessary) condition for simultaneous satisfaction of these two is $\omega_2 h_{\alpha_{a_H}} - \omega_1 h_{\alpha_{a_C}} = (\omega_2-\omega_1)h_{\alpha_v}$; i.e. 
	\begin{equation}
		\dfrac{\alpha_{a_H}\omega_{2}-\omega_{2}+\omega_{1}}{\omega_{1}}=\alpha_{a_C}~,
		\label{eqn:condition 1} 
	\end{equation}
	with $\alpha_{a_C}>0$.
	(ii) In order to have positive work done by the system and also positivity of heat absorbed and heat transferred, impose $\text{Tr}(\delta\rho^{H}h_{\alpha'})>0$ with $\alpha'=\alpha_v$, $\alpha'=\alpha_{a_H}$ and $\alpha'= \alpha_{a_C}$ for these three different quantities. 
	Second condition together with (\ref{B34}) then implies that $(\omega_2-\omega_1)\textrm{Tr}(\delta\rho^H h_{\alpha_v})<\omega_2\textrm{Tr}(\delta\rho^H h_{\alpha_{a_H}})$. Then, since $\eta_0<1$, by (\ref{B35}) we have $\eta_E<1$.
	As $\alpha_{a_C}>0$, \eqref{eqn:condition 1} implies $\eta_{0}<\alpha_{a_H}$. Now, since we want $\eta_E > \eta_0$, then according to earlier discussion one has $\alpha_{a_H}<1$. Therefore $0<\eta_0<1$ implies that for our process we need to choose $\alpha_{a_H}$ such that
	\begin{equation}
		0<\eta_{0}<\alpha_{a_H}<1~.
		\label{eqn:constraint1}
	\end{equation}
	Consequently if we choose $\alpha_{a_H}$ from Eq. \eqref{eqn:constraint1} then from Eq \eqref{eqn:condition 1} we get that,
	\begin{equation}
		0<\alpha_{a_C}<\alpha_{a_H}~.
		\label{N2}
	\end{equation}
	The argument is as follows. Let $\alpha_{a_C}>\alpha_{a_H}$, then (\ref{eqn:condition 1}) implies $\alpha_{a_H}>1$ which violates \eqref{eqn:constraint1}. Hence this is not possible, so we have at most
	$\alpha_{a_C}\leq\alpha_{a_H}$.
	If $\alpha_{a_C}=\alpha_{a_H}$,  then \eqref{eqn:condition 1} implies they must be equal to unity, which is again in contrary to \eqref{eqn:constraint1}. Therefore, we must have (\ref{N2}).
	Let us come back to the second condition again. For the present case one finds
	\begin{equation}
		\text{Tr}(\delta\rho^{H}h_{\alpha'})=\pm2(1+\alpha')I_{1}~.
		\label{eqn:traceinInotation}
	\end{equation}
	Positive sign is for $\ket{s}$ while the negative sign is for $\ket{a}$. In general $I_1$ is a function of $\alpha_{a_H}, \omega_{2}$ and $\tau_a$ (for explicit expressions in $(1+1)$ and $(1+3)$ spacetime dimensions, refer to \cite{App123}).
	Now for a choice of $\alpha_{a_H}, \omega_{2}$, satisfying (\ref{eqn:constraint1}) and (\ref{N2}), we have to choose $\tau_a$, the half time duration of acceleration of first detector during heating, such that the cyclic condition along with condition (ii) is being satisfied. If one chooses $\tau_a$ such that $I_1>0$, then initial state must be $\ket{s}$ in order to satisfy condition (ii). Whereas if $I_1<0$ for a $\tau_a$, then one needs to consider $\ket{\chi}=\ket{a}$. This describes a complete protocol to build a cycle for $\eta_E>\eta_0$ with $\eta_E<1$.

	To summarise, here we proposed a quantum Otto engine using two qubits in the lights of quantum entanglement and relativity. We found that entanglement and relativity together might be useful to get enhanced efficiency. The efficiency becomes double to that of QOE when the qubits initially are in a maximally entangled state, and the frame of measurement is not accelerating during the heating process. If the acceleration of the frame of reference is less than that of the other qubit, then one has $\eta_E>\eta_0$. We also provided a detailed protocol to choose the available parameters in order to obtain enhancement of efficiency. 
		
		This analysis shows a possibility of building a QOE with greater efficiency and in turn may suggest an indirect approach for verifying the Unruh effect. Therefore, if an experimental setup for EUQOE can be built, then by measuring the efficiency, one may investigate the properties of Unruh radiation. Moreover, the dependence of efficiency on the accelerations of the qubits can be helpful to set up an apparatus for indirect experimental verification of the Unruh effect.  The investigation in this direction is kept as a future goal.
	

	\vskip 3mm
	\noindent
	{\bf Acknowledgment:}
	GRK would like to express his gratitude to the Department of Physics, IIT Guwahati, for the addition of the Mini-Project course in the curriculum to promote early research experience.
	GRK and BRM would like to thank Dipankar Barman for reviewing calculations and spotting typos in the first version of the draft.
	The research of BRM is partially supported by a START- UP RESEARCH GRANT (No. SG/PHY/P/BRM/01) from the Indian Institute of Technology Guwahati, India and by a Core Research Grant (File no. CRG/2020/000616) from Science and Engineering Research Board (SERB), Department of Science $\&$ Technology (DST), Government of India. 

	
	
	\newpage
	
	\clearpage
	\begin{widetext}
		\begin{center}	
			{\bf Supplementary material}	
		\end{center}
		\appendix
		\section{Relation between proper times of the two uniformly accelerated frames}\label{App1}
		This analysis is followed from \cite{Rodriguez-Camargo:2016fbq} (see the discussion at the beginning of section 2 of this reference).
		Consider two uniformly accelerated detectors both are moving in the right Rindler wedge with acceleration parameters are given by $a_1$ and $a_2$.  For simplicity we take their motion along Minkowski $X$-axis. The trajectories on the Minkowski spacetime are given by the following relations among the coordinates:
		\begin{eqnarray}
			&&T = \frac{e^{a_1\xi_1}}{a_1}\sinh(a_1\eta_1)~;
			\nonumber
			\\
			&&X = \frac{e^{a_1\xi_1}}{a_1}\cosh(a_1\eta_1)~,
			\label{AA1}
		\end{eqnarray}
		for first detector and 
		\begin{eqnarray}
			&&T=\dfrac{e^{a_{2}\xi_2}}{a_{2}}\sinh(a_2\eta_2)~;
			\nonumber
			\\
			&& X=\dfrac{e^{a_{2}\xi_2}}{a_{2}}\cosh(a_{2}\eta_2)~,
			\label{AA2}
		\end{eqnarray}
		for the second detector.
		These relations in terms of the detector's proper time are given by
		\begin{eqnarray}
			&&T = \frac{1}{A_1}\sinh(A_1\tau_1)~;
			\nonumber
			\\
			&&X = \frac{1}{A_1}\cosh(A_1\tau_1)~,
			\label{AA3}
		\end{eqnarray}
		and 
		\begin{eqnarray}
			&&T=\dfrac{1}{A_2}\sinh(A_2\tau_2)~;
			\nonumber
			\\
			&& X=\dfrac{1}{A_2}\cosh(A_2\tau_2)~,
			\label{AA4}
		\end{eqnarray}
		respectively. In the above $A_1$ and $A_2$ are the proper accelerations while $\tau_1$ and $\tau_2$ are their proper times. These are related to $a_1, a_2$ and $\eta_1, \eta_2$ by the following relations:
		\begin{eqnarray}
			&&A_1 = a_1 e^{-a_1\xi_1}; \,\,\, A_2 = a_2 e^{-a_2\xi_2}~;
			\nonumber
			\\
			&&\tau_1 = \eta_1 e^{a_1\xi_1}; \,\,\, \tau_2 = \eta_2 e^{a_2\xi_2}~.
			\label{AA5}
		\end{eqnarray}
		On a constant $T/X$ line both the detectors satisfy the following relation 
		\begin{equation}
			A_1\tau_1 = A_2\tau_2~.
			\label{AA6}
		\end{equation}
		For the choice $\xi_1=0=\xi_2$ in their respective frames, $a_1, a_2$ are identified as proper accelerations while $\eta_1, \eta_2$ are then their respective proper times. Thus for this simple choice we have the relation between the respective proper times as
		\begin{equation}
			\tau_2 =\alpha_a \tau_1~,
			\label{AA7}
		\end{equation}
		where $\alpha_a = a_1/a_2$.

		For constant velocity phase we can have $\tau_2=\alpha_v\tau_1$, where 
		\begin{equation}
			\alpha_{v}=\sqrt{1-v_{rel}^{2}}
		\end{equation}
		where, $v_{rel}$ is the relative velocity between the detectors. If the detectors are moving with same constant velocity, then $v_{rel}=0$ and hence $\alpha_v=1$.

		
		\section{Time evolution of density operator} \label{App2}
		The evolution of density operatorwill find out in interaction picture.
		In the interaction picture we express the interaction Hamiltonian (\ref{B7}) as
		\begin{equation}
			H_{int}^{I}=\mu\Big(m_{1}^{I}\Phi(X_{1}(\tau_{1}))\dfrac{d\tau_{1}}{d\tau}+m_{2}^{I}\Phi(X_{2}(\tau_{2}))\dfrac{d\tau_{2}}{d\tau}\Big)~,
			\label{B17}
		\end{equation}
		where, $m_{i}^{I}=\mathcal{U}^{i\dagger}m_{i}\mathcal{U}^i$ with $\mathcal{U}^i = \exp(-i \int H_id\tau_i)$. Here $H_i$ denotes part of (\ref{eqn:internalhamiltonian}) which corresponds to $i^{th}$ qubit. Note that when $\rho$ changes, the level spacing $\omega$ remains fixed. This is happening in stage 2 and stage 4. In that case $H_i$ is taken to be time independent and hence $\mathcal{U}^i = \exp(-i H_i\Delta\tau^s_i)$ corresponding to $s^{th}$ stage. Since $m_i$ is given by (\ref{B9}), we reexpress (\ref{B17}) as
		\begin{equation}
			H_{int}^{I}=M_{1}\dfrac{d\tau_{1}}{d\tau}\Phi(X_{1}(\tau_{1}))+M_{2}\dfrac{d\tau_{2}}{d\tau}\Phi(X_{2}(\tau_{2}))~,
			\label{B18}
		\end{equation}
		where
		\begin{equation}
			M_{1}\equiv\mu m_{1}^{I}=\mu\mathcal{U}^{1\dagger}m_{1}\mathcal{U}^1
			=\mu\begin{pmatrix}
				0&0&e^{i\omega\Delta\tau^s_{1}}&0\\
				0&0&0&e^{i\omega\Delta\tau^s_{1}}\\
				e^{-i\omega\Delta\tau^s_{1}}&0&0&0\\
				0&e^{-i\omega\Delta\tau^s_{1}}&0&0\\
			\end{pmatrix}~;
			\label{B19}
		\end{equation}
		and
		\begin{equation}
			M_{2}\equiv\mu m_2^I= \mu\mathcal{U}^{2\dagger}m_{2}\mathcal{U}^2=\mu
			\begin{pmatrix}
				0&e^{i\omega\Delta\tau^s_{2}}&0&0\\
				e^{-i\omega\Delta\tau^s_{2}}&0&0&0\\
				0&0&0&e^{i\omega\Delta\tau^s_{2}}\\
				0&0&e^{-i\omega\Delta\tau^s_{2}}&0\\
			\end{pmatrix}~.
			\label{B20}	
		\end{equation}
		The matrix forms are represented in the basis $\{\ket{e_1,e_2},\ket{e_1,g_2},\ket{g_1,e_2},\ket{g_1,g_2}\}$.

		The initial density matrix for the collective system, composed of two qubits and the scalar field, in interaction picture is given by
		\begin{equation}
			\rho_0^I =\rho_0 = \rho_{A_0}\otimes\rho_{f_0}~,
			\label{B21}
		\end{equation}
		with the initial state of the field is represented by density operator $\rho_{f_0}=\ket{0_{M}}\bra{0_{M}}$ by considering initially the field is in Minkowski vacuum state $\ket{0_M}$. The evolution of density operator is determined by the equation
		\begin{equation}
			i\dfrac{d\rho^{I}(\tau)}{d\tau}=[H^{I}_{int},\rho^{I}(\tau)]~.
			\label{B22}
		\end{equation}
		Here everything has to be measured from the first detector. So the proper time $\tau$ in above has to be chosen as $\tau_1$.

		The solution of the Eq. (\ref{B22}) is normally achieved by well known perturbative approach. Till the second order in perturbation series it is given by
		\begin{equation}
			\rho^{I}(t)=\underbrace{\rho^{I}(t_0)}_{\mathcal{O}(\mu^0)} -\underbrace{i\int_{t_0}^{t}[H^{I}_{int}(t'),\rho^{I}(t_{0})]dt'}_{\mathcal{O}(\mu^1)} - \underbrace{T\int_{t_0}^{t} dt'\int_{t_0}^{t'} dt'' [[H^{I}_{int}(t'),[[H^{I}_{int}(t''),\rho^{I}(t_{0})]]}_{\mathcal{O}_(\mu^2)} +\mathcal{O}(\mu^3)~,
			\label{A1}
		\end{equation}
		where, $T$ means time-ordered. product.
		By removing the time order product, the can also be expressed as
		\begin{equation}
			\rho^{I}(t)=\rho^{I}(t_{0})-i\int_{t_{0}}^{t}[H^{I}_{int}(t'),\rho^{I}(t_{0})]dt'-\dfrac{1}{2}\int_{t_{0}}^{t}\int_{t_{0}}^{t}[H^{I}_{int}(t'),[H^{I}_{int}(t''),\rho^{I}(t_{0})]]dt''dt'~.
			\label{A2}
		\end{equation}
		Here $t$ is the clock time of our observer. For the main purpose, we need to choose $t=\tau_1$, which we will consider later.

		As we are interested in the evolution of the two detector system, the field degrees of freedom must be traced out. At time $t=t_{0}$, the composite state is represented by (\ref{B21}). 
		The integrant in first order term of Eq. (\ref{A2}), after taking trace over all field states, yields
		\begin{eqnarray}
			&&\Trf(([H^{I}_{int}(t'),\rho^{I}(t_{0})])=\sum_{\{\ket{\Theta}\}=\text{set of all field states}}\inner{\Theta}{[H^{I}_{int}(t'),\rho^{I}(t_{0})]}{\Theta}
			\nonumber
			\\
			&&=\sum_{\{\ket{\Theta}\}}\inner{\Theta}{\Big(H_{int}^{I}(t')\rho_{A_0}\otimes\ket{0_{M}}\bra{0_{M}}-\rho_{A_0}\otimes\ket{0_{M}}\bra{0_{M}}H_{int}^{I}(t')\Big)}{\Theta}
			\nonumber
			\\
			&&=\inner{0_{M}}{\Big(\dfrac{d\tau_{1}'}{dt'}M_{1}(\tau_{1}')\Phi(X_{1}(\tau_{1}'))+\dfrac{d\tau_{2}'}{dt'}M_{2}(\tau_{2}')\Phi(X_{2}(\tau_{2}'))\Big)\rho_{A_0}\ket{0_M}
				\nonumber
				\\
				&&-\bra{0_M}\rho_{A_0}\Big(\dfrac{d\tau_{1}'}{dt'}M_{1}(\tau_{1}')\Phi(X_{1}(\tau_{1}'))+\dfrac{d\tau_{2}'}{dt'}M_{2}(\tau_{2}')\Phi(X_{2}(\tau_{2}'))\Big)}{0_{M}}~.
			\label{A3}
		\end{eqnarray}
		In the last step we substituted the explicit expression for interaction Hamiltonian (\ref{B18}).
		Now as $\inner{0_{M}}{\Phi(\tau_{i}')}{0_{M}}=0$, one finds $\Trf(([H^{I}_{int}(t'),\rho^{I}(t_{0})])=0$; i.e. first order term does not contribute to the perturbative solution. Let us now concentrate on the next order term in (\ref{A2}).
		
		The second-order term consists of several parts. For the present discussion, we drop the notation ``$int$'' and will use it only when necessary. The expansion of brackets in the integrant of the $\mathcal{O}(\mu^2)$ terms gives
		\begin{eqnarray}
			[H(\tau'),[H(\tau''),\rho(\tau_{0})]]&=&H(\tau')H(\tau'')\rho(\tau_{0})-H(\tau')\rho(\tau_{0})H(\tau'')
			\nonumber
			\\
			&-&H(\tau'')\rho(\tau_{0})H(\tau')+\rho(\tau_{0})H(\tau'')H(\tau')~.
			\label{A4}
		\end{eqnarray}
		Next using explicit expressions (\ref{B18}) and (\ref{B21}) we have to take trace over all field states. The first term of the above yields
		\begin{equation}
			\begin{split}
				\Trf((H(\tau')H(\tau'')\rho^{I}(\tau_{0}))=\sum_{\{\ket{\Theta}\}}\inner{\Theta}{\Big[\Big(\dfrac{d\tau_{1}'}{dt'}M_{1}(\tau')\Phi(X_{1}(\tau'))+\dfrac{d\tau_{2}'}{dt'}M_{2}(\tau')\Phi(X_{2}(\tau'))\Big)
					\\
					\times\Big(\dfrac{d\tau_{1}''}{dt''}M_{1}(\tau'')\Phi(X_{1}(\tau''))+\dfrac{d\tau_{2}''}{dt''}M_{2}(\tau'')\Phi(X_{2}(\tau''))\Big)\rho_{A_0}\ket{0_M}\bra{0_M}\Big]}{\Theta}
				\\
				=M_{1}(\tau_{1}')M_{1}(\tau_{1}'')\rho_{A_0}\avg{1}{\tau'}{1}{\tau''}\dfrac{d\tau_{1}'}{dt'}\dfrac{d\tau_{1}''}{dt''}+	M_{1}(\tau_{1}')M_{2}(\tau_{2}'')\rho_{A_0}\avg{1}{\tau'}{2}{\tau''}\dfrac{d\tau_{1}'}{dt'}\dfrac{d\tau_{2}''}{dt''}
				\\	
				+M_{2}(\tau_{2}')M_{1}(\tau_{1}'')\rho_{A_0}\avg{2}{\tau'}{1}{\tau''}\dfrac{d\tau_{1}''}{dt''}\dfrac{d\tau_{2}'}{dt'}+	M_{2}(\tau_{2}')M_{2}(\tau_{2}'')\rho_{A_0}\avg{2}{\tau'}{2}{\tau''}\dfrac{d\tau_{2}'}{dt'}\dfrac{d\tau_{2}''}{dt''}~.
			\end{split}
			\label{A5}
		\end{equation}
		In the above we used the notation $\bra{0_M}{\Phi(X_{i}(\tau_{i}'))}{\Phi(X_{j}(\tau_{j}''))}\ket{0_M}=\avg{j}{\tau''}{i}{\tau'}$.
		Similarly, other terms lead to
		\begin{equation}
			\begin{split}
				\Trf((H(\tau')\rho^{I}(\tau_{0}) H(\tau''))=M_{1}(\tau_{1}')\rho_{A_0}M_{1}(\tau_{1}'')\avg{1}{\tau''}{1}{\tau'}\dfrac{d\tau_{1}'}{dt'}\dfrac{d\tau_{1}''}{dt''}
				\\
				+M_{1}(\tau_{1}')\rho_{A_0}M_{2}(\tau_{2}'')\avg{2}{\tau''}{1}{\tau'}\dfrac{d\tau_{1}'}{dt'}\dfrac{d\tau_{2}''}{dt''}
				\\
				+M_{2}(\tau_{2}')\rho_{A_0}M_{1}(\tau_{1}'')\avg{1}{\tau''}{2}{\tau'}\dfrac{d\tau_{1}''}{dt''}\dfrac{d\tau_{2}'}{dt'}+M_{2}(\tau_{2}')\rho_{A0}M_{2}(\tau_{2}'')\avg{2}{\tau''}{2}{\tau'}\dfrac{d\tau_{2}'}{dt'}\dfrac{d\tau_{2}''}{dt''}~;
			\end{split}
			\label{A6}
		\end{equation}
		\begin{equation}
			\begin{split}
				\Trf((H(\tau'')\rho^{I}(\tau_{0}) H(\tau'))=M_{1}(\tau_{1}'')\rho_{A_0}M_{1}(\tau_{1}')\avg{1}{\tau'}{1}{\tau''}\dfrac{d\tau_{1}'}{dt'}\dfrac{d\tau_{1}''}{dt''}
				\\
				+M_{1}(\tau_{1}'')\rho_{A_0}M_{2}(\tau_{2}')\avg{2}{\tau'}{1}{\tau''}\dfrac{d\tau_{1}''}{dt''}\dfrac{d\tau_{2}'}{dt'}
				\\
				+M_{2}(\tau_{2}'')\rho_{A_0}M_{1}(\tau_{1}')\avg{1}{\tau'}{2}{\tau''}\dfrac{d\tau_{1}'}{dt'}\dfrac{d\tau_{2}''}{dt''}+M_{2}(\tau_{2}'')\rho_{A_0}M_{2}(\tau_{2}')\avg{2}{\tau'}{2}{\tau''}\dfrac{d\tau_{2}'}{dt'}\dfrac{d\tau_{2}''}{dt''}~;
			\end{split}
			\label{A7}
		\end{equation}
		and 
		\begin{equation}
			\begin{split}
				\Trf((\rho^{I}(\tau_{0})H(\tau'') H(\tau'))=\rho_{A_0}M_{1}(\tau_{1}'')M_{1}(\tau_{1}')\avg{1}{\tau''}{1}{\tau'}\dfrac{d\tau_{1}'}{dt'}\dfrac{d\tau_{1}''}{dt''}
				\\
				+\rho_{A_0}M_{1}(\tau_{1}'')M_{2}(\tau_{2}')\avg{1}{\tau''}{2}{\tau'}\dfrac{d\tau_{1}''}{dt''}\dfrac{d\tau_{2}'}{dt'}
				\\	
				+\rho_{A_0}M_{2}(\tau_{2}'')M_{1}(\tau_{1}')\avg{2}{\tau''}{1}{\tau'}\dfrac{d\tau_{1}'}{dt'}\dfrac{d\tau_{2}''}{dt''}+\rho_{A_0}M_{2}(\tau_{1}')M_{2}(\tau_{2}'')\avg{2}{\tau''}{2}{\tau'}\dfrac{d\tau_{2}'}{dt'}\dfrac{d\tau_{2}''}{dt''}~.
			\end{split}
			\label{A8}
		\end{equation}
		
		Substitution of these in (\ref{A2}) yields (\ref{B23}) with $\delta\rho$ is given by 
		\begin{equation}
			\rho(\tau^f_1) \equiv \rho^I(\tau^f_{1},\tau^f_{2}(\tau^f_1))=\rho_{A_0}+\delta\rho(\tau^f_{1},\tau^f_{2}(\tau^f_1))~,
			\label{B23}
		\end{equation}
		where
		\begin{eqnarray}
			&&\delta\rho(\tau^f_{1},\tau^f_{2}(\tau^f_1))=-\dfrac{1}{2}\Big[\int_{\tau^i_{1}}^{\tau^f_{1}}\int_{\tau^i_{1}}^{\tau^f_{1}}\Gamma_{11}\avg{1}{\tau'}{1}{\tau''} d\tau_{1}''d\tau_{1}'
			+\int_{\tau^i_{1}}^{\tau^f_{1}}\int_{\tau^i_{2}}^{\tau^f_{2}}\Gamma_{12}^{1}\avg{1}{\tau'}{2}{\tau''} d\tau_{2}''d\tau_{1}'
			\nonumber
			\\
			&&+\int_{\tau^i_{1}}^{\tau^f_{1}}\int_{\tau^i_{2}}^{\tau^f_{2}}\Gamma_{12}^{2}\avg{1}{\tau''}{2}{\tau'} d\tau_{2}'d\tau_{1}''
			+\int_{\tau^i_{1}}^{\tau^f_{1}}\int_{\tau^i_{2}}^{\tau^f_{2}}\Gamma_{21}^{1}\avg{2}{\tau'}{1}{\tau''} d\tau_{2}'d\tau_{1}''
			\nonumber
			\\
			&&+\int_{\tau^i_{1}}^{\tau^f_{1}}\int_{\tau^i_{2}}^{\tau^f_{2}}\Gamma_{21}^{2}\avg{2}{\tau''}{1}{\tau'} d\tau_{2}''d\tau_{1}'
			+\int_{\tau^i_{2}}^{\tau^f_{2}}\int_{\tau^i_{2}}^{\tau^f_{2}}\Gamma_{22}\avg{2}{\tau'}{2}{\tau''} d\tau_{2}''d\tau_{2}'\Big]~.
			\label{B24}		
		\end{eqnarray}
		In the above the final expression will be achieved after replacing $\tau_2$ in terms of $\tau_1$ by $\tau_2=\alpha\tau_1$. 
		The explicit expressions for $\Gamma$'s are given by
		\begin{eqnarray}
			&&\Gamma_{11} = M_{1}(\tau_{1}')M_{1}(\tau_{1}'')\rho_{A_0} - M_{1}(\tau_{1}')\rho_{A0}M_{1}(\tau_{1}'') - M_{1}(\tau_{1}'')\rho_{A_0}M_{1}(\tau_{1}')
			\nonumber
			\\
			&&+\rho_{A_0}M_{1}(\tau_{1}'')M_{1}(\tau_{1}')~;
			\label{A9}
			\\
			&&\Gamma_{12}^{1} = M_{1}(\tau_{1}')M_{2}(\tau_{2}'')\rho_{A_0} - M_{2}(\tau_{2}'')\rho_{A_0}M_{1}(\tau_{1}')~;
			\label{A10}
			\\
			&&\Gamma_{12}^{2} = \rho_{A_0}M_{1}(\tau_{1}'')M_{2}(\tau_{2}') - M_{2}(\tau_{2}')\rho_{A_0}M_{1}(\tau_{1}'') ~;
			\label{A11}
			\\
			&&\Gamma_{21}^{1} = M_{2}(\tau_{2}')M_{1}(\tau_{1}'')\rho_{A_0}- M_{1}(\tau_{1}'')\rho_{A_0}M_{2}(\tau_{2}')~;
			\label{A12}
			\\
			&&\Gamma_{21}^{2} = \rho_{A_0}M_{2}(\tau_{2}'')M_{1}(\tau_{1}')-
			M_{1}(\tau_{1}')\rho_{A_0}M_{2}(\tau_{2}'') ~;
			\label{A13}
			\\
			&&\Gamma_{22} = M_{2}(\tau_{2}')M_{2}(\tau_{2}'')\rho_{A_0} - M_{2}(\tau_{2}')\rho_{A_0}M_{2}(\tau_{2}'') - M_{2}(\tau_{2}'')\rho_{A_0}M_{2}(\tau_{2}')
			\nonumber
			\\
			&&+\rho_{A_0}M_{2}(\tau_{1}')M_{2}(\tau_{2}'')~.
			\label{A14}
		\end{eqnarray}
		
		\section{Matrix representations of the operators appearing in $\Gamma$,s}\label{App3}
		The matrix forms of the operators, appearing in Eq. (\ref{A9}) to Eq. (\ref{A14}), in the basis $\{\ket{e_1,e_2},\ket{e_1,g_2},\ket{g_1,e_2},\ket{g_1,g_2}\}$ can be obtained by using (\ref{B11}), (\ref{B19}) and (\ref{B20}). They are listed below. Below we denote $\Delta\tau_{1}'=\tau_{1}'+\tau_{a}$ and $\Delta\tau_{2}'=\tau_{1}'+\tau_{b}$, where $-\tau_{a}$ and $-\tau_{b}$, the initial proper times of first and second detectors respectively, signify the time durations of interaction with the field. Same goes for $\Delta\tau_{1}'',\Delta\tau_{2}''$ respectively as well. 
		
		\begin{enumerate}
			\item $M_{1}(\tau_{1}')M_{1}(\tau_{1}'')\rho_{A_0}$:
			\begin{equation}
				\begin{pmatrix}
					pe^{i\omega(\Delta\tau_{1}'-\Delta\tau_{1}'')}&0&0&0\\
					0&q\vert b_{1}\vert^{2}e^{i\omega(\Delta\tau_{1}'-\Delta\tau_{1}'')}&qb_{1}b_{2}^{*}e^{i\omega(\Delta\tau_{1}'-\Delta\tau_{1}'')}&0\\
					0&qb_{1}^{*}b_{2}e^{-i\omega(\Delta\tau_{1}'-\Delta\tau_{1}'')}&q\vert b_{2}\vert^{2}e^{-i\omega(\Delta\tau_{1}'-\Delta\tau_{1}'')}&0\\
					0&0&0&0\\
				\end{pmatrix}
				\label{C1}
			\end{equation}
			
			\item $M_{1}(\tau_{1}')M_{2}(\tau_{2}'')\rho_{A_0}$:
			\begin{equation}
				\begin{pmatrix}
					0&0&0&0\\
					0&q b_{1}^{*}b_{2}e^{i\omega(\Delta\tau_{1}'-\Delta\tau_{2}'')}&q\vert b_{2}\vert^{2}e^{i\omega(\Delta\tau_{1}'-\Delta\tau_{2}'')}&0\\
					0&q \vert b_{1}\vert^{2}e^{-i\omega(\Delta\tau_{1}'-\Delta\tau_{2}'')}&qb_{1} b_{2}^{*}e^{-i\omega(\Delta\tau_{1}'-\Delta\tau_{2}'')}&0\\
					pe^{-i\omega(\Delta\tau_{1}'+\Delta\tau_{2}'')}&0&0&0\\
				\end{pmatrix}
				\label{C2}
			\end{equation}
			
			\item $M_{2}(\tau_{2}')M_{1}(\tau_{1}'')\rho_{A_0}$:
			\begin{equation}
				\begin{pmatrix}
					0&0&0&0\\
					0&q b_{1}^{*}b_{2}e^{-i\omega(\Delta\tau_{2}'-\Delta\tau_{1}'')}&q\vert b_{2}\vert^{2}e^{-i\omega(\Delta\tau_{2}'-\Delta\tau_{1}'')}&0\\
					0&q \vert b_{1}\vert^{2}e^{i\omega(\Delta\tau_{2}'-\Delta\tau_{1}'')}&qb_{1} b_{2}^{*}e^{i\omega(\Delta\tau_{2}'-\Delta\tau_{1}'')}&0\\
					pe^{-i\omega(\Delta\tau_{2}'+\Delta\tau_{1}'')}&0&0&0\\
				\end{pmatrix}
				\label{C3}
			\end{equation}
			
			\item $M_{2}(\tau_{2}')M_{2}(\tau_{2}'')\rho_{A_0}$:
			\begin{equation}
				\begin{pmatrix}
					pe^{i\omega(\Delta\tau_{2}'-\Delta\tau_{2}'')}&0&0&0\\
					0&q\vert b_{1}\vert^{2}e^{-i\omega(\Delta\tau_{2}'-\Delta\tau_{2}'')}&qb_{1}b_{2}^{*}e^{-i\omega(\Delta\tau_{2}'-\Delta\tau_{2}'')}&0\\
					0&qb_{1}^{*}b_{2}e^{i\omega(\Delta\tau_{2}'-\Delta\tau_{2}'')}&q\vert b_{2}\vert^{2}e^{i\omega(\Delta\tau_{2}'-\Delta\tau_{2}'')}&0\\
					0&0&0&0\\
				\end{pmatrix}
				\label{C4}
			\end{equation}
			
			\item $\rho_{A_0}M_{1}(\tau_{1}'')M_{1}(\tau_{1}')$:
			\begin{equation}
				\begin{pmatrix}
					pe^{-i\omega(\Delta\tau_{1}'-\Delta\tau_{1}'')}&0&0&0\\
					0&q\vert b_{1}\vert^{2}e^{-i\omega(\Delta\tau_{1}'-\Delta\tau_{1}'')}&qb_{1}b_{2}^{*}e^{i\omega(\Delta\tau_{1}'-\Delta\tau_{1}'')}&0\\
					0&qb_{1}^{*}b_{2}e^{-i\omega(\Delta\tau_{1}'-\Delta\tau_{1}'')}&q\vert b_{2}\vert^{2}e^{i\omega(\Delta\tau_{1}'-\Delta\tau_{1}'')}&0\\
					0&0&0&0\\
				\end{pmatrix}
				\label{C5}
			\end{equation}
			
			\item $\rho_{A_0}M_{1}(\tau_{1}'')M_{2}(\tau_{2}')$:
			\begin{equation}
				\begin{pmatrix}
					0&0&0&pe^{i\omega(\Delta\tau_{1}''+\Delta\tau_{2}')}\\
					0&qb_{1}b_{2}^{*}e^{-i\omega(\Delta\tau_{1}''-\Delta\tau_{2}')}&q\vert b_{1}\vert^{2}e^{i\omega(\Delta\tau_{1}''-\Delta\tau_{2}')}&0\\
					0&q\vert b_{2}\vert^{2}e^{-i\omega(\Delta\tau_{1}''-\Delta\tau_{2}')}&qb_{1}^{*}b_{2}e^{i\omega(\Delta\tau_{1}''-\Delta\tau_{2}')}&0\\
					0&0&0&0\\
				\end{pmatrix}
				\label{C6}
			\end{equation}
			
			\item $\rho_{A_0}M_{2}(\tau_{2}'')M_{1}(\tau_{1}')$:
			\begin{equation}
				\begin{pmatrix}
					0&0&0&pe^{i\omega(\Delta\tau_{2}''+\Delta\tau_{1}')}\\
					0&qb_{1}b_{2}^{*}e^{i\omega(\Delta\tau_{2}''-\Delta\tau_{1}')}&q\vert b_{1}\vert^{2}e^{-i\omega(\Delta\tau_{2}''-\Delta\tau_{1}')}&0\\
					0&q\vert b_{2}\vert^{2}e^{i\omega(\Delta\tau_{2}''-\Delta\tau_{1}')}&qb_{1}^{*}b_{2}e^{-i\omega(\Delta\tau_{2}''-\Delta\tau_{1}')}&0\\
					0&0&0&0\\
				\end{pmatrix}
				\label{C7}
			\end{equation}
			
			\item $\rho_{A_0}M_{2}(\tau_{2}'')M_{2}(\tau_{2}')$:
			\begin{equation}
				\begin{pmatrix}
					pe^{i\omega(\Delta\tau_{2}''-\Delta\tau_{2}')}&0&0&0\\
					0&q\vert b_{1}\vert^{2}e^{-i\omega(\Delta\tau_{2}''-\Delta\tau_{2}')}&qb_{1}b_{2}^{*}e^{i\omega(\Delta\tau_{2}''-\Delta\tau_{2}')}&0\\
					0&qb_{1}^{*}b_{2}e^{-i\omega(\Delta\tau_{2}''-\Delta\tau_{2}')}&q\vert b_{2}\vert^{2}e^{i\omega(\Delta\tau_{2}''-\Delta\tau_{2}')}&0\\
					0&0&0&0\\
				\end{pmatrix}
				\label{C8}
			\end{equation}
			
			\item $M_{1}(\tau_{1}')\rho_{A_0}M_{1}(\tau_{1}'')$:
			\begin{equation}
				\begin{pmatrix}
					q\vert b_{2}\vert^{2}e^{i\omega(\Delta\tau_{1}'-\Delta\tau_{1}'')}&0&0&qb_{1}^{*}b_{2}e^{i\omega(\Delta\tau_{1}'+\Delta\tau_{1}'')}\\
					0&0&0&0\\
					0&0&pe^{-i\omega(\Delta\tau_{1}'-\Delta\tau_{1}'')}&0\\
					qb_{1}b_{2}^{*}e^{-i\omega(\Delta\tau_{1}'+\Delta\tau_{1}'')}&0&0&q\vert b_{1}\vert^{2}e^{-i\omega(\Delta\tau_{1}'-\Delta\tau_{1}'')}\\
				\end{pmatrix}
				\label{C9}
			\end{equation}
			
			\item $M_{1}(\tau_{1}')\rho_{A_0}M_{2}(\tau_{2}'')$:
			\begin{equation}
				\begin{pmatrix}
					qb_{1}^{*}b_{2}e^{i\omega(\Delta\tau_{1}'-\Delta\tau_{2}'')}&0&0&q\vert b_{2}\vert^{2}e^{i\omega(\Delta\tau_{1}'+\Delta\tau_{2}'')}\\
					0&0&0&0\\
					0&pe^{-i\omega(\Delta\tau_{1}'-\Delta\tau_{2}'')}&0&0\\
					q\vert b_{1}\vert^{2}e^{-i\omega(\Delta\tau_{1}'+\Delta\tau_{2}'')}&0&0&qb_{1}b_{2}^{*}e^{-i\omega(\Delta\tau_{1}'-\Delta\tau_{2}'')}\\
				\end{pmatrix}
				\label{C10}
			\end{equation}
			
			\item $M_{2}(\tau_{2}')\rho_{A_0}M_{1}(\tau_{1}'')$:
			\begin{equation}
				\begin{pmatrix}
					qb_{1}b_{2}^{*}e^{i\omega(\Delta\tau_{2}'-\Delta\tau_{1}'')}&0&0&q\vert b_{1}\vert^{2}e^{i\omega(\Delta\tau_{2}'+\Delta\tau_{1}'')}\\
					0&0&pe^{-i\omega(\Delta\tau_{2}'-\Delta\tau_{1}'')}&0\\
					0&0&0&0\\
					q\vert b_{2}\vert^{2}e^{-i\omega(\Delta\tau_{2}'+\Delta\tau_{1}'')}&0&0&qb_{1}^{*}b_{2}e^{-i\omega(\Delta\tau_{2}'-\Delta\tau_{1}'')}\\
				\end{pmatrix}
				\label{C11}
			\end{equation}
			
			\item $M_{2}(\tau_{2}')\rho_{A_0}M_{2}(\tau_{2}'')$:
			\begin{equation}
				\begin{pmatrix}
					q\vert b_{1}\vert^{2}e^{i\omega(\Delta\tau_{2}'-\Delta\tau_{2}'')}&0&0&qb_{1}b_{2}^{*}e^{i\omega(\Delta\tau_{2}'+\Delta\tau_{2}'')}\\
					0&pe^{-i\omega(\Delta\tau_{2}'-\Delta\tau_{2}'')}&0&0\\
					0&0&0&0\\
					qb_{1}^{*}b_{2}e^{-i\omega(\Delta\tau_{2}'+\Delta\tau_{2}'')}&0&0&q\vert b_{2}\vert^{2}e^{-i\omega(\Delta\tau_{2}'-\Delta\tau_{2}'')}\\
				\end{pmatrix}
				\label{C12}
			\end{equation}
		\end{enumerate}
		We use them in Appendix \ref{App5} to evaluate Eq. (\ref{B37}).
		
		\section{Structure of Wightman function}\label{App4}
		Here we give the explicit forms of the Wightman functions, defined as
		\begin{equation}
			\avg{i}{\tau'}{j}{\tau''}=G_{ij}(\tau_{i}',\tau_{j}'')~,
			\label{D1}	
		\end{equation}
		for different situations as appeared in Eq. (\ref{B24}). Also we will find few properties among them which will be used in next Appendix. Here the expression in $(1+1)$ dimensions will be given. This is sufficient as similar properties also hold in $(1+3)$ dimensions. We provide the forms which are considered in \cite{Barman:2021oum} as we will see later that such is efficient to tackle the situation elegantly. For real massless scalar field, Wightman function, with respect to Rindler frame with $\ket{0_M}$ as vacuum, is obtained by taking $\beta\rightarrow\infty$ of those in \cite{Barman:2021oum}. This is given by  
		\begin{eqnarray}
			G_{jl}(\Delta\xi,\Delta\eta)&=&\int_{-\infty}^{\infty}\dfrac{dk}{8\pi\omega_{k}\sqrt{\sinh\Big(\frac{\pi\omega_{k}}{a_{j}}\Big)\sinh\Big(\frac{\pi\omega_{k}}{a_{l}}\Big)}}\Big[e^{ik\Delta\xi-i\omega_{k}\Delta\eta}e^{\frac{\pi\omega_{k}}{2}(\frac{1}{a_{j}}+\frac{1}{a_{l}})}
			\nonumber
			\\
			&+& e^{ik\Delta\xi+i\omega_{k}\Delta\eta}e^{-\frac{\pi\omega_{k}}{2}(\frac{1}{a_{j}}+\frac{1}{a_{l}})}\Big]~.
			\label{D2}
		\end{eqnarray}
		This expression is already in Rindler coordinates. Using the relation between the proper times of the detectors $\tau_2 = \alpha\tau_1$, we express the above one in terms of proper time of first detector. The Wightman functions, appearing in (\ref{B24}), come out to be in the following forms. 
		\begin{enumerate}
			
			\item $G_{12}(\tau_{1}',\tau_{2}'')$: $\Delta\xi=0,\Delta\eta=\tau_{1}'-\alpha_a\tau_{1}''$
			\begin{eqnarray}
				G_{12}(\tau_{1}',\tau_{2}''(\tau_{1}'')) &=& \int_{-\infty}^{\infty}\dfrac{dk}{8\pi\omega_{k}\sqrt{\sinh\Big(\frac{\pi\omega_{k}}{a_{1}}\Big)\sinh\Big(\frac{\pi\omega_{k}}{a_{2}}\Big)}}\Big[e^{-i\omega_{k}(\tau_{1}'-\alpha_{a}\tau_{1}'')}e^{\frac{\pi\omega_{k}}{2a_{1}}(1+\alpha_{a})}
				\nonumber
				\\		
				&+& e^{+i\omega_{k}(\tau_{1}'-\alpha_{a}\tau_{1}'')}e^{-\frac{\pi\omega_{k}}{2a_{1}}(1+\alpha_{a})}\Big]~;
			\end{eqnarray}
			
			\item $G_{12}(\tau_{1}'',\tau_{2}'(\tau_{1}'))$: $\Delta\xi=0,\Delta\eta=\tau_{1}''-\alpha_{a}\tau_{1}'$
			\begin{eqnarray}
				G_{12}(\tau_{1}'',\tau_{2}'(\tau_{1}')) &=& \int_{-\infty}^{\infty}\dfrac{dk}{8\pi\omega_{k}\sqrt{\sinh\Big(\frac{\pi\omega_{k}}{a_{1}}\Big)\sinh\Big(\frac{\pi\omega_{k}}{a_{2}}\Big)}}\Big[e^{-i\omega_{k}(\tau_{1}''-\alpha_{a}\tau_{1}')}e^{\frac{\pi\omega_{k}}{2a_{1}}(1+\alpha_{a})}
				\nonumber
				\\
				&+& e^{+i\omega_{k}(\tau_{1}''-\alpha_{a}\tau_{1}')}e^{-\frac{\pi\omega_{k}}{2a_{1}}(1+\alpha_{a})}\Big]~;
				\label{N1}
			\end{eqnarray}
			
			\item $G_{21}(\tau_{2}'(\tau_{1}'),\tau_{1}'')$: $\Delta\xi=0,\Delta\eta=\alpha_{a}\tau_{1}'-\tau_{1}''$
			\begin{eqnarray}
				G_{21}(\tau_{2}'(\tau_{1}'),\tau_{1}'') &=& \int_{-\infty}^{\infty}\dfrac{dk}{8\pi\omega_{k}\sqrt{\sinh\Big(\frac{\pi\omega_{k}}{a_{1}}\Big)\sinh\Big(\frac{\pi\omega_{k}}{a_{2}}\Big)}}\Big[e^{-i\omega_{k}(\alpha_{a}\tau_{1}'-\tau_{1}'')}e^{\frac{\pi\omega_{k}}{2a_{1}}(1+\alpha_{a})}
				\nonumber
				\\
				&+& e^{+i\omega_{k}(\alpha_{a}\tau_{1}'-\tau_{1}'')}e^{-\frac{\pi\omega_{k}}{2a_{1}}(1+\alpha_{a})}\Big]~;
			\end{eqnarray}
			
			\item $G_{21}(\tau_{2}''(\tau_{1}''),\tau_{1}')$: $\Delta\xi=0,\Delta\eta=\alpha_{a}\tau_{1}''-\tau_{1}'$
			\begin{eqnarray}
				G_{21}(\tau_{2}''(\tau_{1}''),\tau_{1}') &=& \int_{-\infty}^{\infty}\dfrac{dk}{8\pi\omega_{k}\sqrt{\sinh\Big(\frac{\pi\omega_{k}}{a_{1}}\Big)\sinh\Big(\frac{\pi\omega_{k}}{a_{2}}\Big)}}\Big[e^{-i\omega_{k}(\alpha_{a}\tau_{1}''-\tau_{1}')}e^{\frac{\pi\omega_{k}}{2a_{1}}(1+\alpha_{a})}
				\nonumber
				\\
				&+& e^{+i\omega_{k}(\alpha_{a}\tau_{1}''-\tau_{1}')}e^{-\frac{\pi\omega_{k}}{2a_{1}}(1+\alpha_{a})}\Big]~.
			\end{eqnarray}
		\end{enumerate}
		
		In the above $\omega_{k}=\vert k\vert$ and hence we can rewrite them in the following forms as well:
		\begin{eqnarray}
			&&G_{12}(\tau_{1}',\tau_{2}''(\tau_{1}''))=\int_{-\infty}^{0}\dfrac{dk}{8\pi(-k)\sqrt{\sinh\Big(\frac{\pi k}{a_{1}}\Big)\sinh\Big(\frac{\pi k}{a_{2}}\Big)}}\Big[e^{ik(\tau_{1}'-\alpha_{a}\tau_{1}'')}e^{\frac{-\pi k}{2a_{1}}(1+\alpha_{a})}+e^{-ik(\tau_{1}'-\alpha_{a}\tau_{1}'')}e^{\frac{\pi k}{2a_{1}}(1+\alpha_{a})}\Big]
			\nonumber
			\\
			&+& \int^{\infty}_{0}\dfrac{dk}{8\pi k\sqrt{\sinh\Big(\frac{\pi k}{a_{1}}\Big)\sinh\Big(\frac{\pi k}{a_{2}}\Big)}}\Big[e^{-ik(\tau_{1}'-\alpha_{a}\tau_{1}'')}e^{\frac{\pi k}{2a_{1}}(1+\alpha_{a})}+e^{ik(\tau_{1}'-\alpha_{a}\tau_{1}'')}e^{-\frac{\pi k}{2a_{1}}(1+\alpha_{a})}\Big]
			\nonumber
			\\
			&=& \int_{\infty}^{0}\dfrac{d(-k)}{8\pi k\sqrt{\sinh\Big(\frac{\pi k}{a_{1}}\Big)\sinh\Big(\frac{\pi k}{a_{2}}\Big)}}\Big[e^{-ik(\tau_{1}'-\alpha_{a}\tau_{1}'')}e^{\frac{\pi k}{2a_{1}}(1+\alpha_{a})}+e^{ik(\tau_{1}'-\alpha_{a}\tau_{1}'')}e^{-\frac{\pi k}{2a_{1}}(1+\alpha_{a})}\Big]
			\nonumber
			\\
			&+&\int^{\infty}_{0}\dfrac{dk}{8\pi k\sqrt{\sinh\Big(\frac{\pi k}{a_{1}}\Big)\sinh\Big(\frac{\pi k}{a_{2}}\Big)}}\Big[e^{-ik(\tau_{1}'-\alpha_{a}\tau_{1}'')}e^{\frac{\pi k}{2a_{1}}(1+\alpha_{a})}+e^{ik(\tau_{1}'-\alpha_{a}\tau_{1}'')}e^{-\frac{\pi k}{2a_{1}}(1+\alpha_{a})}\Big]
			\nonumber
			\\
			&=& \int^{\infty}_{0}\dfrac{dk}{4\pi k\sqrt{\sinh\Big(\frac{\pi k}{a_{1}}\Big)\sinh\Big(\frac{\pi k}{a_{2}}\Big)}}\Big[e^{-ik(\tau_{1}'-\alpha_{a}\tau_{1}'')}e^{\frac{\pi k}{2a_{1}}(1+\alpha_{a})}+e^{ik(\tau_{1}'-\alpha_{a}\tau_{1}'')}e^{-\frac{\pi k}{2a_{1}}(1+\alpha_{a})}\Big]~.
		\end{eqnarray}
		Similarly, others can be expressed as
		\begin{equation}
			G_{12}(\tau_{1}'',\tau_{2}'(\tau_{1}'))=
			\int^{\infty}_{0}\dfrac{dk}{4\pi k\sqrt{\sinh\Big(\frac{\pi k}{a_{1}}\Big)\sinh\Big(\frac{\pi k}{a_{2}}\Big)}}\Big[e^{-ik(\tau_{1}''-\alpha_{a}\tau_{1}')}e^{\frac{\pi k}{2a_{1}}(1+\alpha_{a})}+e^{ik(\tau_{1}''-\alpha_{a}\tau_{1}')}e^{-\frac{\pi k}{2a_{1}}(1+\alpha_{a})}\Big]~;
		\end{equation}
		\begin{equation}
			G_{21}(\tau_{2}'(\tau_{1}'),\tau_{1}'')=
			\int^{\infty}_{0}\dfrac{dk}{4\pi k\sqrt{\sinh\Big(\frac{\pi k}{a_{1}}\Big)\sinh\Big(\frac{\pi k}{a_{2}}\Big)}}\Big[e^{-ik(\alpha_{a}\tau_{1}'-\tau_{1}'')}e^{\frac{\pi k}{2a_{1}}(1+\alpha_{a})}+e^{ik(\alpha_{a}\tau_{1}'-\tau_{1}'')}e^{-\frac{\pi k}{2a_{1}}(1+\alpha_{a})}\Big]~;
		\end{equation}
		and
		\begin{equation}
			G_{21}(\tau_{2}''(\tau_{1}''),\tau_{1}')=
			\int^{\infty}_{0}\dfrac{dk}{4\pi k\sqrt{\sinh\Big(\frac{\pi k}{a_{1}}\Big)\sinh\Big(\frac{\pi k}{a_{2}}\Big)}}\Big[e^{-ik(\alpha_{a}\tau_{1}''-\tau_{1}')}e^{\frac{\pi k}{2a_{1}}(1+\alpha_{a})}+e^{ik(\alpha_{a}\tau_{1}''-\tau_{1}')}e^{-\frac{\pi k}{2a_{1}}(1+\alpha_{a})}\Big]~.
		\end{equation}
		These explicit expressions shows the following property:
		\begin{equation}
			G_{12}(\tau_{1}',\tau_{2}''(\tau_{1}''))=G_{21}(-\tau_{2}''(\tau_{1}''),-\tau_{1}')~,
			\label{eqn:wightmanrelation}
		\end{equation}
		which will be used in the next Appendix.
		
		We now look at the structure of $G_{11}(\tau_{1}',\tau_{1}'')$ and $G(\tau_{2}'(\tau_{1}'),\tau_{2}''(\tau_{1}''))$. Like the earlier way we find 
		\begin{equation}
			G_{11}(\tau_{1}',\tau_{1}'')=\int^{\infty}_{0}\dfrac{dk}{4\pi k\sinh\Big(\frac{\pi k}{a_{1}}\Big)}\Big[e^{-ik(\tau_{1}'-\tau_{1}'')}e^{\frac{\pi k}{a_{1}}}+e^{ik(\tau_{1}'-\tau_{1}'')}e^{-\frac{\pi k}{a_{1}}}\Big]~;
			\label{eqn:G11}
		\end{equation}
		and 
		\begin{equation}
			G_{22}(\tau_{2}'(\tau_{1}'),\tau_{1}''(\tau_{1}''))=\int^{\infty}_{0}\dfrac{dk}{4\pi k\sinh\Big(\frac{\pi k}{a_{2}}\Big)}\Big[e^{-ik\alpha_{a}(\tau_{1}'-\tau_{1}'')}e^{\frac{\pi k}{a_{2}}}+e^{ik\alpha_{a}(\tau_{1}'-\tau_{1}'')}e^{-\frac{\pi k}{a_{2}}}\Big]~.
		\end{equation}
		Using $\dfrac{1}{a_{2}}=\dfrac{\alpha_{a}}{a_{1}}$ and changing the variable $\alpha_{a} k=y$ we find
		\begin{equation}
			G_{22}(\tau_{2}'(\tau_{1}'),\tau_{1}''(\tau_{1}''))=\int^{\infty}_{0}\dfrac{dy}{4\pi y\sinh\Big(\frac{\pi y}{a_{1}}\Big)}\Big[e^{-iy(\tau_{1}'-\tau_{1}'')}e^{\frac{\pi y}{a_{1}}}+e^{iy(\tau_{1}'-\tau_{1}'')}e^{-\frac{\pi y}{a_{1}}}\Big]~.
		\end{equation}
		Hence, we get a relation
		\begin{equation}
			G_{22}(\tau_{2}'(\tau_{1}'),\tau_{2}''(\tau_{1}''))=G_{11}(\tau_{1}',\tau_{1}'')~.
			\label{eqn:wightmanrelation1}
		\end{equation}
		This property is also essential for the next Appendix. Same properties are also applicable to the (1+3) dimension structure of the Wightman function. The expression for (1+3) dimension structure as given in \cite{Barman:2021oum} is,
		\begin{equation}
			\begin{split}
				G_{jl}(\Delta\eta_{jl})=\int_{0}^{\infty}d\omega_{k}\int\dfrac{d^{2}k_{p}}{(2\pi)^{4}}\dfrac{2}{\sqrt{a_{j}a_{l}}}\mathcal{K}\Big[\dfrac{i\omega_{k}}{a_{j}},\dfrac{\vert k_{p}\vert}{a_{j}}\Big]\mathcal{K}\Big[\dfrac{i\omega_{k}}{a_{l}},\dfrac{\vert k_{p}\vert}{a_{l}}\Big]\\
				\times[e^{-i\omega_{k}\eta_{jl}}e^{\frac{\pi\omega_{k}}{2}(\frac{1}{a_{j}}+\frac{1}{a_{l}})}+e^{i\omega_{k}\eta_{jl}}e^{-\frac{\pi\omega_{k}}{2}(\frac{1}{a_{j}}+\frac{1}{a_{l}})}]
			\end{split}
			\label{eqn:1+3}
		\end{equation}
		where, $\mathcal{K}[n, z]$ denotes the modified Bessel function of
		the second kind of order $n$. This expression will also be used in following Appendices.

		\section{Derivation of Eq. (\ref{B39})}\label{App5}
		
		To calculate the efficiency, we have to take the trace of product between $\delta\rho_{A}^{\alpha}$ and $h_{\alpha'}$. As $h_{\alpha'}$ is time-independent (refer to Eq. \eqref{B25}), while taking the product, we can take it inside the integral of Eq. \eqref{B24}. There is a total of six terms corresponding to each $\Gamma$, and every term contains factors like $\textrm{Tr}(\Gamma h_{\alpha})$ in the integrant. We will calculate these traces one by one. The matrix forms of $\Gamma$,s can be evaluated by using results of Appendix \ref{App3} in equations (\ref{A9}) to $(\ref{A14})$. We list below the values of the traces of each term in (\ref{B24}).
		In the following equations, $-\tau_{a}$ is the initial time of first detector when the interaction is turned on.
		\begin{enumerate}
			\item Trace with $\Gamma_{12}^{1}$
			\begin{equation}
				2q\mu^{2}\alpha_{a}[b_{1}^{*}b_{2}e^{i\omega((\tau_{1}'+\tau_{a})-\alpha_{a}(\tau_{1}''+\tau_{a}))}-b_{1}b_{2}^{*}e^{-i\omega((\tau_{1}'+\tau_{a})-\alpha_{a}(\tau_{1}''+\tau_{a}))}]~;
				\label{eqn:g1212}
			\end{equation}
			
			\item Trace with $\Gamma_{12}^{2}$
			\begin{equation}
				2q\mu^{2}\alpha'\alpha_{a}[b_{1}^{*}b_{2}e^{-i\omega(\alpha_{a}(\tau_{1}'+\tau_{a})-(\tau_{1}''+\tau_{a}))}-b_{1}b_{2}^{*}e^{i\omega(\alpha_{a}(\tau_{1}'+\tau_{a})-(\tau_{1}''+\tau_{a}))}]~;
				\label{eqn:g1222}
			\end{equation}
			
			\item Trace with $\Gamma_{21}^{1}$
			\begin{equation}
				2q\mu^{2}\alpha'\alpha_{a}[b_{1}b_{2}^{*}e^{i\omega(\alpha_{a}(\tau_{1}'+\tau_{a})-(\tau_{1}''+\tau_{a}))}-b_{1}^{*}b_{2}e^{-i\omega(\alpha_{a}(\tau_{1}'+\tau_{a})-(\tau_{1}''+\tau_{a}))}]~;
				\label{eqn:g2112}
			\end{equation}
			
			\item Trace with $\Gamma_{21}^{2}$
			\begin{equation}
				2q\mu^{2}\alpha_{a}[b_{1}b_{2}^{*}e^{-i\omega((\tau_{1}'+\tau_{a})-\alpha_{a}(\tau_{1}''+\tau_{a}))}-b^{*}_{1}b_{2}e^{i\omega((\tau_{1}'+\tau_{a})-\alpha_{a}(\tau_{1}''+\tau_{a}))}]~;
				\label{eqn:g2122}
			\end{equation}
			
			\item Trace with $\Gamma_{11}$
			\begin{equation}
				4\mu^{2}\cos(\omega(\tau_{1}'-\tau_{1}''))[p+q(\vert b_{1}\vert^{2}-\vert b_{2}\vert^{2})]~.
				\label{E1}
			\end{equation}
			
			\item Trace with $\Gamma_{22}$
			\begin{equation}
				4\mu^{2}\alpha'\alpha_{a}^{2}\cos(\omega\alpha_{a}(\tau_{1}'-\tau_{1}''))[p+q(\vert b_{2}\vert^{2}-\vert b_{1}\vert^{2})]~.
				\label{E2}
			\end{equation}
		\end{enumerate}
		
		Once we have the list of all the traces, we multiply the traces with their corresponding Wightman functions as appeared in Eq. \eqref{B24}. We consider the interaction to be on within the time range  $[-\tau_{a},\tau_{a}]$, measured by first detector's clock. The reason for this has been explained above Eq. (\ref{B14}) and the value is given by Eq. (\ref{B14}). We will use the property of the Wightman function, represented by Eq. \eqref{eqn:wightmanrelation} to simplify equations. We will simplify the expression pairwise.  First we take terms corresponding to \eqref{eqn:g1222} and \eqref{eqn:g2112}. These two lead to
		\begin{equation}
			\begin{split}
				\int_{-\tau_{a}}^{\tau_{a}}\int_{-\tau_{a}}^{\tau_{a}}\Big[2q\mu^{2}\alpha'\alpha_{a}[b_{1}^{*}b_{2}e^{-i\omega(\alpha_{a}(\tau_{1}'+\tau_{a})-(\tau_{1}''+\tau_{a}))}-b_{1}b_{2}^{*}e^{i\omega(\alpha_{a}(\tau_{1}'+\tau_{a})-(\tau''_{1}+\tau_{a}))}]G_{12}(\tau_{1}'',\tau_{2}'(\tau_{1}'))\\+2q\mu^{2}\alpha'\alpha_{a}[b_{1}b_{2}^{*}e^{i\omega(\alpha_{a}(\tau_{1}'+\tau_{a})-(\tau_{1}''+\tau_{a}))}-b_{1}^{*}b_{2}e^{-i\omega(\alpha_{a}(\tau_{1}'+\tau_{a})-(\tau_{1}''+\tau_{a}))}]G_{21}(\tau_{2}'(\tau_{1}'),\tau_{1}'')\Big]d\tau_{1}''d\tau_{1}'~.
			\end{split}
			\label{E3}	
		\end{equation}
		We change the variables from $\tau_{1}'\rightarrow -\tau_{1}',\tau_{1}''\rightarrow -\tau_{1}''$ in the second term corresponding to $G_{21}(\tau_{2}'(\tau_{1}'),\tau_{1}'')$. This gives us,
		\begin{equation}
			\begin{split}
				2q\mu^{2}\alpha'\alpha_{a}\int_{-\tau_{a}}^{\tau_{a}}\int_{-\tau_{a}}^{\tau_{a}}\Big[[b_{1}^{*}b_{2}e^{-i\omega(\alpha_{a}(\tau_{1}'+\tau_{a})-(\tau_{1}''+\tau_{a}))}-b_{1}b_{2}^{*}e^{i\omega(\alpha_{a}(\tau_{1}'+\tau_{a})-(\tau_{1}''+\tau_{a}))}]G_{12}(\tau_{1}'',\tau_{2}'(\tau_{1}'))\\+[b_{1}b_{2}^{*}e^{-i\omega(\alpha_{a}(\tau_{1}'-\tau_{a})-(\tau_{1}''-\tau_{a}))}-b_{1}^{*}b_{2}e^{i\omega(\alpha_{a}(\tau_{1}'-\tau_{a})-(\tau_{1}''-\tau_{a}))}]G_{21}(-\tau_{2}'(\tau_{1}'),-\tau_{1}'')\Big]d\tau_{1}''d\tau_{1}'\\
				= 2q\mu^{2}\alpha'\alpha_{a}\int_{-\tau_{a}}^{\tau_{a}}\int_{-\tau_{a}}^{\tau_{a}}\Big[b_{1}b_{2}^{*}[e^{-i\omega(\alpha_{a}(\tau_{1}'-\tau_{a})-(\tau_{1}''-\tau_{a}))}-e^{i\omega(\alpha_{a}(\tau_{1}'+\tau_{a})-(\tau_{1}''+\tau_{a}))}]\\+b_{1}^{*}b_{2}[e^{-i\omega(\alpha_{a}(\tau_{1}'+\tau_{a})-(\tau_{1}''+\tau_{a}))}-e^{i\omega(\alpha_{a}(\tau_{1}'-\tau_{a})-(\tau_{1}''-\tau_{a}))}]\Big]G_{12}(\tau_{1}'',\tau_{2}'(\tau_{1}'))d\tau_{1}''d\tau_{1}'~.
			\end{split}
			\label{E4}	
		\end{equation}
		Rearranging the terms finally we get,
		\begin{equation}
			\begin{split}
				-4iq\mu^{2}\alpha'\alpha_{a}\int_{-\tau_{a}}^{\tau_{a}}\int_{-\tau_{a}}^{\tau_{a}}\Big[b_{1}b_{2}^{*}e^{i\omega(\alpha_{a}-1)\tau_{a}}+b_{1}^{*}b_{2}e^{-i\omega(\alpha_{a}-1)\tau_{a}}\Big]\sin(\omega(\alpha_{a}\tau_{1}'-\tau_{1}''))G_{12}(\tau_{1}'',\tau_{2}'(\tau_{1}'))d\tau_{1}''d\tau_{1}'~.
				\label{eqn:GG12}
			\end{split}
		\end{equation}
		
		Next we concentrate on terms corresponding to \eqref{eqn:g1212} and\eqref{eqn:g2122}. They collectively yield
		\begin{equation}
			\begin{split}
				\int_{-\tau_{a}}^{\tau_{a}}\int_{-\tau_{a}}^{\tau_{a}}\Big[2q\mu^{2}\alpha_{a}[b_{1}^{*}b_{2}e^{i\omega((\tau_{1}'+\tau_{a})-\alpha_{a}(\tau_{1}''+\tau_{a}))}-b_{1}b_{2}^{*}e^{-i\omega((\tau_{1}'+\tau_{a})-\alpha_{a}(\tau_{1}''+\tau_{a}))}]G_{12}(\tau_{1}',\tau_{2}''(\tau_{1}''))\\+2q\mu^{2}\alpha_{a}[b_{1}b_{2}^{*}e^{-i\omega((\tau_{1}'+\tau_{a})-\alpha_{a}(\tau_{1}''+\tau_{a}))}-b^{*}_{1}b_{2}e^{i\omega((\tau_{1}'+\tau_{a})-\alpha_{a}(\tau_{1}''+\tau_{a}))}]G_{21}(\tau_{2}''(\tau_{1}''),\tau_{1}')\Big]d\tau_{1}''d\tau_{1}'~.
			\end{split}
			\label{E5}
		\end{equation}
		In similar way the above simplifies to
		\begin{equation}
			-4iq\mu^{2}\alpha_{a}	\int_{-\tau_{a}}^{\tau_{a}}\int_{-\tau_{a}}^{\tau_{a}}\Big[b_{1}b_{2}^{*}e^{i\omega(\alpha_{a}-1)\tau_{a}}+b^{*}_{1}b_{2}e^{-i\omega(\alpha_{a}-1)\tau_{a}}\Big]\sin(\omega(\tau_{1}'-\alpha_{a}\tau_{1}''))G_{21}(\tau_{2}''(\tau_{1}''),\tau_{1}')d\tau_{1}''d\tau_{1}'~.
			\label{eqn:GG21}
		\end{equation}
		Before going to other two terms, we now add terms \eqref{eqn:GG12} and \eqref{eqn:GG21}. This leads to, upon using $\tau'_1\to -\tau'_1$ and $\tau''_1\to -\tau''_1$,
		\begin{equation}
			\begin{split}
				-4i\alpha'q\mu^{2}\alpha_{a}\int_{-\tau_{a}}^{\tau_{a}}\int_{-\tau_{a}}^{\tau_{a}}\Big[b_{1}b_{2}^{*}e^{i\omega(\alpha_{a}-1)\tau_{a}}+b_{1}^{*}b_{2}e^{-i\omega(\alpha_{a}-1)\tau_{a}}\Big]\sin(\omega(\alpha_{a}\tau_{1}'-\tau_{1}''))G_{12}(\tau_{1}'',\tau_{2}'(\tau_{1}'))d\tau_{1}''d\tau_{1}'\\
				-4iq\mu^{2}\alpha	\int_{-\tau_{a}}^{\tau_{a}}\int_{-\tau_{a}}^{\tau_{a}}\Big[b_{1}b_{2}^{*}e^{i\omega(\alpha_{a}-1)\tau_{a}}+b^{*}_{1}b_{2}e^{-i\omega(\alpha_{a}-1)\tau_{a}}\Big]\sin(\omega(\alpha_{a}\tau_{1}''-\tau_{1}'))G_{21}(-\tau_{2}''(\tau_{1}''),-\tau_{1}')d\tau_{1}''d\tau_{1}'~.		
			\end{split}
			\label{E6}
		\end{equation}
		Next using (\ref{eqn:wightmanrelation}) and as $\tau_{1}',\tau_{1}''$ are integration variables, we can interchange them in the second expression. Therefore whole expression simplifies to
		\begin{equation}
			-4iq\mu^{2}\alpha_{a}(1+\alpha')\Big[b_{1}b_{2}^{*}e^{i\omega(\alpha_{a}-1)\tau_{a}}+b_{1}^{*}b_{2}e^{-i\omega(\alpha_{a}-1)\tau_{a}}\Big]\int_{-\tau_{a}}^{\tau_{a}}\int_{-\tau_{a}}^{\tau_{a}}\sin(\omega(\alpha_{a}\tau_{1}'-\tau_{1}''))G_{12}(\tau_{1}'',\tau_{2}'(\tau_{1}'))d\tau_{1}''d\tau_{1}'~.
			\label{eqn:trace1}
		\end{equation} 
		
		Now we will give attention to other two terms, containing (\ref{E1}) and (\ref{E2}). Adding them and then using Eq. \eqref{eqn:wightmanrelation1}, we obtain,
		\begin{eqnarray}
			\int_{-\tau_{a}}^{\tau_{a}}\int_{-\tau_{a}}^{\tau_{a}}4\mu^{2}[\cos(\omega(\tau_{1}'-\tau_{1}''))(p+q(\vert b_{1}\vert^{2}-\vert b_{2}\vert^{2}))
			\nonumber
			\\
			+\alpha'\alpha_{a}^{2}\cos(\omega\alpha_{a}(\tau_{1}'-\tau_{1}''))(p+q(\vert b_{2}\vert^{2}-\vert b_{1}\vert^{2}))]G_{11}(\tau_{1}',\tau_{1}'')d\tau_{1}''d\tau_{1}'~.
			\label{eqn:trace2}
		\end{eqnarray}
		Finally, accumulating Eq. \eqref{eqn:trace1} and Eq. \eqref{eqn:trace2} we obtain our desire expression (\ref{B37}).
		\begin{eqnarray}
			\text{Tr}(\delta\rho^{\alpha} h_{\alpha'}) &=& \int_{-\tau_a}^{\tau_a}\int_{-\tau_a}^{\tau_a}\Big[2iq\mu^{2}\alpha(1+\alpha')[b_{1}b_{2}^{*}e^{i\omega(\alpha-1)\tau_{a}}+b_{1}^{*}b_{2}e^{-i\omega(\alpha-1)\tau_{a}}]
			\nonumber
			\\
			&&\times \sin(\omega(\alpha\tau_{1}'-\tau_{1}''))G_{12}(\tau_{1}'',\tau_{2}'(\tau_{1}'))
			\nonumber
			\\
			&&-2\mu^{2}[\cos(\omega(\tau_{1}'-\tau_{1}''))(p+q(\vert b_{1}\vert^{2}-\vert b_{2}\vert^{2}))+
			\nonumber
			\\
			&&\alpha'\alpha^{2}\cos(\omega\alpha(\tau_{1}'-\tau_{1}''))(p+q(\vert b_{2}\vert^{2}-\vert b_{1}\vert^{2}))]G_{11}(\tau_{1}',\tau_{1}'')\Big]d\tau_{1}''d\tau_{1}'~.
			\label{B37}
		\end{eqnarray}
		In the above we denote $G_{ij} = \langle\Phi_i(\tau_i')\Phi_j(\tau_{j}'')\rangle$ Use of this in (\ref{B35}) in principle gives us the explicit expression for efficiency of our Otto engine. Without evaluation of the integration one can not tell anything more about its properties.  
		
		At this junction, let us look at a special case where $p=0$ and hence $q=1$. It means, according to (\ref{B11}), our system is initially in the entangled state $\ket{\chi}$. In this situation (\ref{B37}) simplifies to
		\begin{eqnarray}
			\text{Tr}(\delta\rho^{\alpha} h_{\alpha'}) &=&\int_{-\tau_a}^{\tau_a}\int_{-\tau_a}^{\tau_a}\Big[2i\mu^{2}\alpha(1+\alpha')[b_{1}b_{2}^{*}e^{i\omega(\alpha-1)\tau_{a}}+b_{1}^{*}b_{2}e^{-i\omega(\alpha-1)\tau_{a}}]
			\nonumber
			\\
			&&\times \sin(\omega(\alpha\tau_{1}'-\tau_{1}''))G_{12}(\tau_{1}'',\tau_{2}'(\tau_{1}'))
			\nonumber
			\\
			&&-2\mu^{2}[\cos(\omega(\tau_{1}'-\tau_{1}''))(\vert b_{1}\vert^{2}-\vert b_{2}\vert^{2})
			\nonumber
			\\
			&&+\alpha'\alpha^{2}\cos(\omega\alpha(\tau_{1}'-\tau_{1}''))(\vert b_{2}\vert^{2}-\vert b_{1}\vert^{2})]G_{11}(\tau_{1}',\tau_{1}'')\Big]d\tau_{1}''d\tau_{1}'~.
			\label{B38}
		\end{eqnarray}
		Now if $\ket{\chi}=\ket{s}$, i.e. the initial state is chosen to be the symmetric entangled state, then we have $b_1=b_{2}=\dfrac{1}{\sqrt{2}}$. This gives us a much more simpler result (\ref{B39}).

		\section{Explicit evaluation of $\textrm{Tr}(\delta\rho^\alpha h_{\alpha'})$}\label{App12}
		$\text{Tr}(\delta\rho^\alpha h_{\alpha'})$ is given by (\ref{B37}).
		We approach the simplification stepwise. First we integrate the term corresponding to $G_{12}(\tau_{1}'',\tau_{2}'(\tau_{1}'))$. Using (\ref{N1}) one finds 
		\begin{eqnarray}
			&&2i\int_{-\tau_{a}}^{\tau_{a}}\int_{-\tau_{a}}^{\tau_{a}}\sin(\omega(\alpha\tau_{1}'-\tau_{1}''))G_{12}(\tau_{1}'',\tau_{2}'(\tau_{1}'))d\tau_{1}''d\tau_{1}'
			\nonumber
			\\
			&&=\int_{0}^{\infty}\dfrac{dk}{4\pi kf(k)}\int_{-\tau_{a}}^{\tau_{a}}\int_{-\tau_{a}}^{\tau_{a}}[e^{i\omega(\alpha\tau_{1}'-\tau_{1}'')}-e^{-i\omega(\alpha\tau_{1}'-\tau_{1}'')}]
			\nonumber
			\\
			&&\times [e^{ik(\alpha\tau_{1}'-\tau_{1}'')}e^{\frac{\pi k(1+\alpha)}{2a_{1}}}+e^{-ik(\alpha\tau_{1}'-\tau_{1}'')}e^{-\frac{\pi k(1+\alpha)}{2a_{1}}}]d\tau_{1}''d\tau_{1}'	~,
		\end{eqnarray}
		where, $f(k)=\sqrt{\sinh\Big(\dfrac{k\pi}{a_{1}}\Big)\sinh\Big(\dfrac{k\pi}{a_{2}}\Big)}$. After integrating over $\tau''_1$ variable and then on $\tau'_1$ and also incorporating the pre-factor in (\ref{B37}) we find term corresponding to $G_{12}(\tau_{1}'',\tau_{2}'(\tau_{1}'))$ as
		\begin{equation}
			\begin{split}
				\dfrac{2\mu^{2}q(1+\alpha')}{\pi}[b_{1}b_{2}^{*}e^{i\omega(\alpha-1)\tau_{a}}+b_{1}^{*}b_{2}e^{-i\omega(\alpha-1)\tau_{a}}]\int_{0}^{\infty}\dfrac{dk\sinh\Big(\dfrac{\pi k(1+\alpha)}{2a_{1}}\Big)}{k\sqrt{\sinh\Big(\dfrac{\pi k}{a_{1}}\Big)\sinh\Big(\dfrac{\pi k\alpha}{a_{1}}\Big)}}\\
				\times \Big[\dfrac{\sin((k+\omega)\tau_{a})\sin(\alpha(k+\omega)\tau_{a})}{(k+\omega)^{2}}-\dfrac{\sin((k-\omega)\tau_{a})\sin(\alpha(k-\omega)\tau_{a})}{(k-\omega)^{2}}\Big]~.
			\end{split}
		\end{equation}
		Now, we simplify the terms corresponding to $G_{11}(\tau_{1}',\tau_{1}'')$. Proceeding in the similar way one finds
		the total term corresponding to $G_{11}(\tau_{1}',\tau_{1}'')$ as
		\begin{equation}
			\begin{split}
				\dfrac{2\mu^{2}}{\pi}\int_{0}^{\infty}\dfrac{dk}{k}\coth\Big(\dfrac{\pi k}{a_{1}}\Big)\Big\{(p+q(\vert b_{1}\vert^{2}-\vert b_{2}\vert^{2}))\Big[\dfrac{\sin^{2}((k+\omega)\tau_{a})}{(k+\omega)^{2}}+\dfrac{\sin^{2}((\omega-k)\tau_{a})}{(\omega-k)^{2}}\Big]\\+\alpha^{2}\alpha'(p+q(\vert b_{2}\vert^{2}-\vert b_{1}\vert^{2}))\Big[\dfrac{\sin^{2}((k+\alpha\omega)\tau_{a})}{(k+\alpha\omega)^{2}}+\dfrac{\sin^{2}((\alpha\omega-k)\tau_{a})}{(\alpha\omega-k)^{2}}\Big]\Big\}~.
			\end{split}
		\end{equation}
		Finally collecting all these terms we find
		\begin{equation}
			\begin{split}
				\text{Tr}(\delta\rho^{\alpha}h_{\alpha'})=\dfrac{2\mu^{2}q(1+\alpha')}{\pi}[b_{1}b_{2}^{*}e^{i\omega(\alpha-1)\tau_{a}}+b_{1}^{*}b_{2}e^{-i\omega(\alpha-1)\tau_{a}}]\int_{0}^{\infty}\dfrac{dk\sinh\Big(\dfrac{\pi k(1+\alpha)}{2a_{1}}\Big)}{k\sqrt{\sinh\Big(\dfrac{\pi k}{a_{1}}\Big)\sinh\Big(\dfrac{\pi k\alpha}{a_{1}}\Big)}}\\
				\times \Big[\dfrac{\sin((k+\omega)\tau_{a})\sin(\alpha(k+\omega)\tau_{a})}{(k+\omega)^{2}}-\dfrac{\sin((k-\omega)\tau_{a})\sin(\alpha(k-\omega)\tau_{a})}{(k-\omega)^{2}}\Big]\\
				-\dfrac{2\mu^{2}}{\pi}\int_{0}^{\infty}\dfrac{dk}{k}\coth\Big(\dfrac{\pi k}{a_{1}}\Big)\Big\{(p+q(\vert b_{1}\vert^{2}-\vert b_{2}\vert^{2}))\Big[\dfrac{\sin^{2}((k+\omega)\tau_{a})}{(k+\omega)^{2}}+\dfrac{\sin^{2}((\omega-k)\tau_{a})}{(\omega-k)^{2}}\Big]\\+\alpha^{2}\alpha'(p+q(\vert b_{2}\vert^{2}-\vert b_{1}\vert^{2}))\Big[\dfrac{\sin^{2}((k+\alpha\omega)\tau_{a})}{(k+\alpha\omega)^{2}}+\dfrac{\sin^{2}((\alpha\omega-k)\tau_{a})}{(\alpha\omega-k)^{2}}\Big]\Big\}~.
				\label{F4}
			\end{split}
		\end{equation}
		A similar calculation can be done for the (1+3) dimension structure of the Wightman's function. It can be noticed that the dependence of Wightman's function in both (1+1) and (1+3) dimensions on $\tau',\tau''$ is identical. Hence using the expression given in Eq. \eqref{eqn:1+3} we get that for (1+3) dimension structure,
		\begin{equation}
			\begin{split}
				\text{Tr}(\delta\rho^{\alpha}h_{\alpha'})=	\dfrac{q\mu^{2}(1+\alpha')}{\pi^{4}\sqrt{a_{1}a_{2}}}[b_{1}b_{2}^{*}e^{i\omega(\alpha-1)\tau_{a}}+b_{1}^{*}b_{2}e^{-i\omega(\alpha-1)\tau_{a}}]\int_{0}^{\infty}d\omega_{k}\int d^{2}k_{p}\mathcal{K}\Big[\dfrac{i\omega_{k}}{a_{1}},\dfrac{\vert k_{p}\vert}{a_{1}}\Big]\mathcal{K}\Big[\dfrac{i\omega_{k}}{a_{2}},\dfrac{\vert k_{p}\vert}{a_{2}}\Big]\\
				\times\sinh\Big(\dfrac{\pi \omega_{k}(1+\alpha)}{2a_{1}}\Big)\Big[\dfrac{\sin((\omega_{k}+\omega)\tau_{a})\sin(\alpha(\omega_{k}+\omega)\tau_{a})}{(\omega_{k}+\omega)^{2}}-\dfrac{\sin((
					\omega_{k}-\omega)\tau_{a})\sin(\alpha(\omega_{k}-\omega)\tau_{a})}{(\omega_{k}-\omega)^{2}}\Big]\\
				-	\dfrac{\mu^{2}}{a_{1}\pi^{4}}\int_{0}^{\infty}d\omega_{k}\int d^{2}k_{p}\mathcal{K}\Big[\dfrac{i\omega_{k}}{a_{1}},\dfrac{\vert k_{p}\vert}{a_{1}}\Big]\mathcal{K}\Big[\dfrac{i\omega_{k}}{a_{1}},\dfrac{\vert k_{p}\vert}{a_{1}}\Big]
				\\\times \cosh\Big(\dfrac{\pi\omega_{k}}{a_{1}}\Big)\Big\{(p+q(\vert b_{1}\vert^{2}-\vert b_{2}\vert^{2}))\Big[\dfrac{\sin^{2}((\omega_{k}+\omega)\tau_{a})}{(\omega_{k}+\omega)^{2}}+\dfrac{\sin^{2}((\omega-\omega_{k})\tau_{a})}{(\omega-\omega_{k})^{2}}\Big]\\+\alpha^{2}\alpha'(p+q(\vert b_{2}\vert^{2}-\vert b_{1}\vert^{2}))\Big[\dfrac{\sin^{2}((\omega_{k}+\alpha\omega)\tau_{a})}{(\omega_{k}+\alpha\omega)^{2}}+\dfrac{\sin^{2}((\alpha\omega-\omega_{k})\tau_{a})}{(\alpha\omega-\omega_{k})^{2}}\Big]\Big\}
			\end{split}
			\label{F5}
		\end{equation}
		Now as in our case $|b_1|=1/\sqrt{2}, |b_2|=1/\sqrt{2}$ and $p=0, q=1$, then taking $\alpha=\alpha_{a_H}$, $\omega=\omega_2$, $a_1=a_{H_1}$ and $a_2 = a_{H_2} = a_{H_1}/\alpha_{a_H}$ in (\ref{F4}) for (1+1) dimensions and (\ref{F5}) for (1+3) dimensions respectively, one finds 
		\begin{equation}
			\text{Tr}(\delta\rho^{H}h_{\alpha'})=\pm2(1+\alpha')I_{1}~,
			\label{eqn:traceinInotation}
		\end{equation}
		where for (1+1) dimensions,
		\begin{equation}
			\begin{split}
				I_{1}=\dfrac{\mu^{2}}{\pi} \cos(\omega_{2}(\alpha_{a_H}-1)\tau_{a})\int_{0}^{\infty}\dfrac{dk\sinh\Big(\dfrac{\pi k(1+\alpha_{a_H})}{2a_{H_1}}\Big)}{k\sqrt{\sinh\Big(\dfrac{\pi k}{a_{H_1}}\Big)\sinh\Big(\dfrac{\pi k\alpha_{a_H}}{a_{H_1}}\Big)}}\\
				\times \Big[\dfrac{\sin((k+\omega_2)\tau_{a})\sin(\alpha_{a_H}(k+\omega_2)\tau_{a})}{(k+\omega_2)^{2}}-\dfrac{\sin((k-\omega_2)\tau_{a})\sin(\alpha_{a_H}(k-\omega_2)\tau_{a})}{(k-\omega_2)^{2}}\Big] \equiv \dfrac{\mu^{2}}{\pi} I~.
			\end{split}
		\end{equation}
		and for (1+3) dimensions,
		\begin{equation}
			\begin{split}
				I_{1}=\dfrac{\mu^{2}\sqrt{\alpha_{a_{H}}}}{2\pi^{4}a_{H_{1}}}\cos(\omega_{2}(\alpha_{a_{H}}-1)\tau_{a})\int_{0}^{\infty}d\omega_{k}\int d^{2}k_{p}\mathcal{K}\Big[\dfrac{i\omega_{k}}{a_{H_{1}}},\dfrac{\vert k_{p}\vert}{a_{H_{1}}}\Big]\mathcal{K}\Big[\dfrac{i\alpha_{a_{H}}\omega_{k}}{a_{H_{1}}},\dfrac{\alpha_{a_{H}}\vert k_{p}\vert}{a_{H_{1}}}\Big]\\
				\times\sinh\Big(\dfrac{\pi k(1+\alpha_{a_{H}})}{2a_{H_{1}}}\Big)\Big[\dfrac{\sin((\omega_{k}+\omega_{2})\tau_{a})\sin(\alpha_{a_{H}}(\omega_{k}+\omega_{2})\tau_{a})}{(\omega_{k}+\omega_{2})^{2}}\\-\dfrac{\sin((\omega_{k}-\omega_{2})\tau_{a})\sin(\alpha_{a_{H}}(\omega_{k}-\omega_{2})\tau_{a})}{(\omega_{k}-\omega_{2})^{2}}\Big]=\dfrac{\mu^{2}\sqrt{\alpha_{a_{H}}}}{2\pi^{4}a_{H_{1}}}I~.
			\end{split}
		\end{equation}
		In the above equation, positive sign is for $\ket{s}$ while negative sign is for $\ket{a}$.

	\end{widetext}

	
	
\end{document}